\documentclass[a4paper,11pt]{article}
\usepackage{jinstpub} 
\usepackage{lineno}
\usepackage{url}
\usepackage{comment}
\usepackage{hyperref}
\usepackage{amsmath}
\usepackage{siunitx}
\usepackage{subcaption}
\usepackage{multirow}
\usepackage[hang,flushmargin]{footmisc}

\newcommand{\blue}[1]{\textcolor{blue}{#1}}

\title{Design, construction, and initial operation of a 30-ton Water-based Liquid Scintillator detector at Brookhaven National Laboratory}

\author[1]{S.~Andrade,}
\affiliation[1]{Brookhaven National Laboratory, Upton, NY 11980, USA}
\author[2]{A.~Baldoni,}
\affiliation[2]{Physics Department, The Pennsylvania State University, State College, PA 16801, USA}
\author[2]{D.F.~Cowen,} 

\author[1]{R.~Diaz Prerez,}

\author[1]{M.V.~Diwan,}

\author[1]{S.~Gokhale,} 

\author[3]{S.~Gwon,}
\affiliation[3]{Department of Physics, Chung-Ang University, Seoul 06974, Republic of Korea}

\author[1, 4]{S.~Hans,} 
\affiliation[4]{Bronx Community College, Bronx, NY 10453, USA}

\author[9]{P. Hackspacher,}

\author[1]{J.~Jerome,}

\author[5]{G.~Lawley,} 
\affiliation[5]{Physics Department, Stony Brook University, Stony Brook, NY 11794, USA}

\author[6,7]{G.D.~Orebi~Gann,} 
\affiliation[6]{Physics Department, University of California at Berkeley, Berkeley, CA 94720-7300, USA}
\affiliation[7]{Lawrence Berkeley National Laboratory, 1 Cyclotron Road, Berkeley, CA 94720-8153, USA}
\author[8]{P.~Kumar,} 
\affiliation[8]{Department of Physics and Astronomy, University of Alabama, Tuscaloosa, AL 35487, USA}

\author[3]{J.~Park,}

\author[1]{C.~Reyes,} 

\author[1]{R.~Rosero,}

\author[1]{N.~Seberg,}
\author[3]{K.~Siyeon,}

\author[6,7]{M.~Smiley,} 

\author[9]{R.~Svoboda,}
\affiliation[9]{Department of Physics and Astronomy, University of California at Davis, Davis, CA 95616, USA}

\author[1]{N.~Speece-Moyer,}

\author[10]{M.~Vagins,}
\affiliation[10]{
Department of Physics and Astronomy
, University of California, Irvine, Irvine, CA 92697, USA}

\author[1]{B.~Walsh,}
\author[8, \dagger]{J.J.~Wang, \note[$\dagger$]{Corresponding author, jwang214@ua.edu}} 

\author[11]{M.~Wilking,}
\affiliation[11]{School of Physics and Astronomy, University of Minnesota, Minneapolis, MN  55455, USA}
\author[1,*]{G.~Yang, \note[*]{Corresponding author, gyang1@bnl.gov}}
\author[9]{D. Wooley,}
\author[1]{M.~Yeh,}

\abstract{

Water-based Liquid Scintillator (WbLS) was proposed over a decade ago as a novel detector medium that might allow the separation and tuning of the relative ratio of the Cherenkov and scintillation signals. A detector employing this technology could support large-scale neutrino detection over both the GeV and MeV energy regimes, while its metal-loading capability could provide an effective means of neutron tagging.
WbLS is attractive for two reasons. It can be deployed in very large detectors. It also allows in-situ tuning of the scintillator concentration, and hence the ratio of Cherenkov to scintillation light. Neither pure water Cherenkov detectors nor conventional liquid scintillator detectors offer this capability.
At Brookhaven National Laboratory (BNL), two prototypes have been built for understanding  WbLS properties and stability, with masses of 1-ton and 30-ton, respectively. We present here the 30-ton prototype detector design, installation, and 
initial operation including the real-time observation of a transient optical response during the staged scintillator injection.
Results from the analysis of data collected in the two detectors will follow in separate publications. 

}


\begin{document}

\maketitle
\flushbottom

\section{Introduction} 
\label{sec:intro}

Water-based Liquid Scintillator (WbLS) technology~\cite{YEH201151} represents a significant advancement in neutrino detection, offering a versatile and potentially cost-effective detection medium for the next generation of large-scale experiments, particularly because water constitutes its primary component. WbLS is created by forming stable, nanometer-scale micelles of liquid scintillator within an aqueous solution. This method seeks to combine the advantages of water Cherenkov~\cite{FUKUDA2003418,BELLERIVE201630} and liquid scintillator detectors~\cite{BOREXINO:2023ygs,PhysRevLett.90.021802,JUNO:2025gmd}, allowing the simultaneous detection and separation of Cherenkov and scintillation light. Several ongoing studies of this medium~\cite{Steiger_2024,Bhattacharya2026BUTTON30,D0MA00055H, Askins2020THEIA} have demonstrated the potential to achieve improved energy resolution compared to a pure water Cherenkov detector, while also offering the possibility of event imaging and particle identification capabilities. These capabilities are important for background rejection and event reconstruction. 
At the percent-level loadings used here, the light yield is below that of a pure liquid scintillator detector~\cite{BOREXINO:2023ygs,PhysRevLett.90.021802}. The WbLS strength is the in-situ tunability of the Cherenkov-to-scintillation ratio in a water-compatible, metal-loadable medium.
Realizing their full potential in a large-scale detector will require addressing key challenges, including achieving the radiopurity levels needed for low-rate measurements such as solar neutrino detection~\cite{DayaBay:2018yms,BELLERIVE201630} and developing event reconstruction techniques in the intermediate optical regime relevant to GeV-scale long-baseline oscillation experiments~\cite{PATTERSON2013151,DUNE:2021tad}. Detailed assessments of the projected physics reach, spanning MeV-scale solar and supernova neutrinos~\cite{Hyper-Kamiokande:2021frf} to GeV-scale oscillation measurements, can be found in Ref.~\cite{Theia}

The promise of this technology has led to its adoption in several experiments. These initiatives are part of a global, phased R\&D program. The goal is to scale WbLS from benchtop concepts to kiloton-scale detectors. Experiments such as ANNIE \cite{ANNIE} at Fermilab utilize WbLS for enhanced neutrino detection, while the \textsc{Eos} detector \cite{EoS} at UC Berkeley has been constructed as a 20-ton performance demonstrator to validate key technologies. The proposed \textsc{Theia} experiment \cite{Theia} is a multi-kiloton detector with a broad physics program, including studies of neutrino oscillations, neutrinoless double beta decay, and solar neutrinos. WbLS is one candidate for the target medium in \textsc{Theia}. The success of these future projects depends on understanding the WbLS behavior at scale.


The R\&D program at Brookhaven National Laboratory (BNL) is central to this global effort. Our 1-ton prototype~\cite{1ton-paper} has been continuously operating for more than two years, during which we demonstrated the fundamental principles of gadolinium (Gd)-compatible WbLS and established the initial stability, light yield, and optical properties of the material~\cite{gwon2025measurementlightyieldresponse}. The WbLS formulation~\footnote{The specific scintillator component concentrations are proprietary and subject to intellectual-property protection and are therefore not disclosed.} used in the present study is identical to that employed in the 1-ton prototype. It consists of a mix of modified surfactants, a fluor of 2,5-diphenyloxazole (PPO)\footnote{CAS Number: 92-71-7}, and a DIN-based (di-isopropylnaphthalen)\footnote{CAS Number: 34640-62-9}, all of which are commercially available. In addition, Gd sulfate has been used previously in Cherenkov detectors such as Super-Kamiokande~\cite{koshio2026upgradesuperkamiokandegadolinium}, and therefore it was selected for the 30-ton tank. Introducing water into a scintillator medium involves several trade-offs, including reduced scintillation light yield, absorption of ultraviolet Cherenkov photons by the scintillator component, and the need for careful radiopurity control through dedicated material screening and radioassay. Unlike traditional liquid scintillator detectors, where water extraction can be followed by phase separation, water is an integral component of WbLS. These effects were investigated previously using the BNL 1T WbLS prototype~\cite{1ton-paper,gwon2025measurementlightyieldresponse}, motivating the further development and scale-up to the 30T demonstrator presented in this work. Building upon these earlier results, the present work focuses on addressing the technical challenges associated with scaling the detector to a substantially larger volume. Demonstrating stable operation at this scale requires addressing several key issues, including monitoring the stability of the liquid over extended operational periods, maintaining a uniform optical response, and implementing reliable purification and circulation systems for large-scale operation.

To address these questions, we designed, constructed, and commissioned a 30-ton WbLS prototype at BNL. This new detector is an order of magnitude larger in mass, serving as an essential bridge between the 1-ton proof-of-concept and future kiloton-scale experiments. The 30-ton system will provide valuable data on long-term stability, optical clarity, and metal-loading capabilities. Its operation provides essential input for de-risking and optimizing the designs of future detectors such as \textsc{Theia}~\cite{Askins2020THEIA}.

This paper presents the technical design, construction, and operational experience of the 30-ton WbLS prototype. We describe the detector design, including the photomultiplier tube (PMT) system, data acquisition, as well as the WbLS production and circulation infrastructure. This includes the nanofiltration (NF) and Gd-loading systems, which have been installed but are not yet commissioned and are planned for future upgrades. We then describe the calibration systems, followed by the detector commissioning and WbLS injection process, including initial detector performance results. The successful operation of the 30 ton demonstrator represents an important milestone toward establishing the viability and scalability of WbLS technology for future large-scale physics experiments.

\section{Detector Design} 
\label{sec:detetcor}

\begin{figure}
    \centering
    \includegraphics[width=1.0\linewidth]{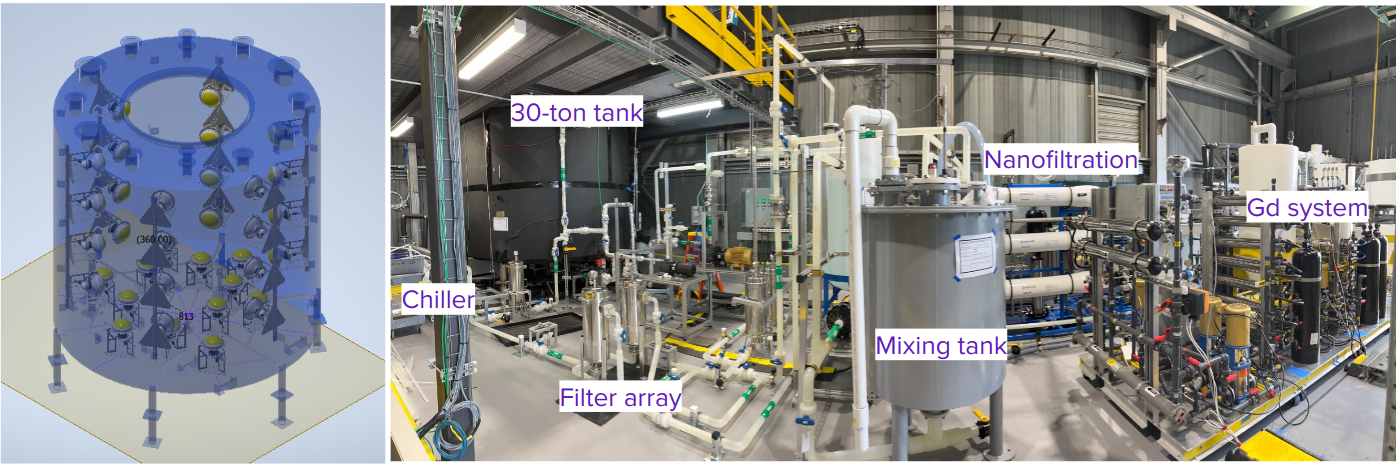}
    \caption{Left: the 30-ton PMT arrangement in the tank. Right: The overview of the 30-ton detector facility. The nanofiltration and Gd systems are planned to be commissioned in future runs.}
    \label{fig:30t_design}
\end{figure}

At the heart of the detector is a cylindrical 
316L stainless steel tank with a radius of 1625.6 mm and a half-height of 1503.35 mm, designed to hold 30 tons of the WbLS medium. Unlike many similar detectors, this tank does not have an internal liner; the WbLS is in direct contact with the stainless steel walls. The inner surface of the tank has been passivated as an engineering measure to improve corrosion resistance and chemical compatibility with the liquid. The compatibility of passivated stainless steel with liquid-scintillator media has been investigated previously, including studies of Gd-loaded liquid scintillator stored in passivated stainless steel~\cite{hino2019aging}.
This design choice motivates material compatibility studies to assess the long-term stability of the detector medium. The entire structure is housed in a dedicated facility that includes advanced circulation and purification systems. Fig.~\ref{fig:30t_design} shows the detector design sketch and the circulation system outlook. Note that the nanofiltration and Gd systems have been tested but have not yet been commissioned; they are planned for future runs.

The detector is instrumented with 36 10-inch Hamamatsu R16367 photomultiplier tubes (PMTs)\footnote{\blue{Hamamatsu Photonics K.K., Hamamatsu City, Japan, \url{https://www.hamamatsu.com}}}, which were specifically selected for their compatibility with WbLS. These PMTs are submerged directly in the detector liquid. The arrangement of the PMTs is designed to optimize light collection and event reconstruction:
\begin{itemize}
    \item Bottom PMTs: Twelve PMTs are arranged in a spiral pattern on the detector floor (Fig. \ref{fig:pmt_location}). This configuration allows for a detailed mapping of light collection efficiency at various radial distances from the center, which is expected to improve the accuracy of event reconstruction.
    \item Side (wall) PMTs: Twenty-four PMTs are mounted in four rows along the cylindrical wall of the tank (Fig. \ref{fig:pmt_location}).
\end{itemize}

This geometric arrangement is crucial for distinguishing between Cherenkov and scintillation light. For vertically traveling cosmic muons, the Cherenkov light is primarily contained within a cone that illuminates the bottom PMTs and the lower two rows of side PMTs (the "in-ring" region). The upper two rows of side PMTs (the "out-ring" region) are largely outside this cone and therefore detect mostly isotropic scintillation light. The difference in the light detected between these two regions provides a method for measuring the scintillation light yield of the WbLS. This PMT geometry provides a photocathode coverage of 7.3\% on the bottom and 7.9\% on the side for an isotropic source at the center of the detector. In addition, two layers of plastic scintillator paddles are installed above and below the tank to tag crossing muons. Each layer consists of four paddles, each measuring $20~\mathrm{cm} \times 90~\mathrm{cm}$. Together, the four paddles provide an active coverage of $80~\mathrm{cm} \times 90~\mathrm{cm}$ at both the top and bottom of the tank.

The detector utilizes a dual trigger system to capture a wide range of physics events. Events can be triggered by:
\begin{itemize}
    \item Cosmic muons passing through a set of plastic scintillator paddles located on top of the detector.
    \item A radioactive source tagging PMT for calibration purposes.
\end{itemize}
This dual-trigger approach allows for the study of both external events (cosmic muons) and internal calibration sources. The data acquisition (DAQ) system (details in Sec. \ref{sec:daq}) is designed to handle the signals from all 36 PMTs, digitizing them for offline analysis.

A key feature of the 30-ton detector is its multiple in-situ circulation and purification systems. The WbLS is continuously circulated during data taking to maintain its optical clarity and stability. This system is equipped with a self-developed, sequential exchange array, a nanofiltration system, and a Gd purification system. The nanofiltration and Gd systems have been tested but have not yet been commissioned. The ability to load the WbLS with metals like gadolinium enhances the detector's capability to tag neutrons, which is important for many neutrino interaction studies, as well as for serving as an active shielding component in direct dark matter detection experiments. Continuous purification is implemented to maintain the optical quality of the detector medium by removing potential contaminants, including suspended particulates, dissolved metal ions or corrosion products, organic impurities or degradation products, and microorganisms such as bacteria. Such contaminants may absorb or scatter scintillation and Cherenkov light, thereby degrading the detector response. Bacterial contamination is a known consideration in large-scale water-based optical detectors, for which UV treatment and microfiltration have been employed to preserve optical quality~\cite{marti2020egads,fukuda2003superk}.
\subsection{PMT} 
\label{sec:pmt}
\begin{figure}[htbp]
  \centering
  \begin{subfigure}[t]{0.48\linewidth}
    \centering
    \includegraphics[width=\linewidth]{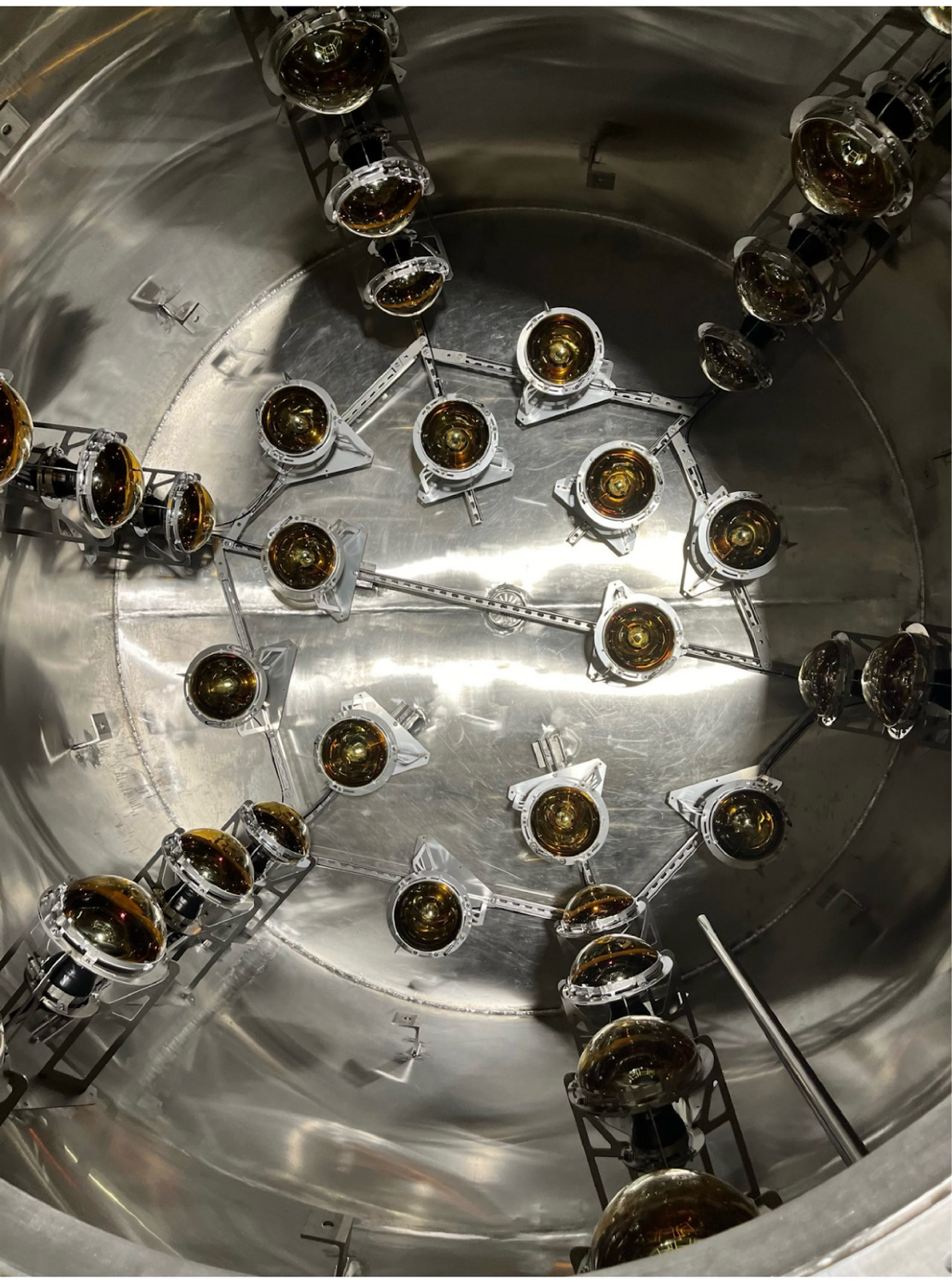}
    \caption{A photograph showing the arrangement of PMTs inside the detector tank. The 12 bottom PMTs are installed in a spiral pattern, while the 24 side PMTs are mounted in four distinct rows along the cylindrical wall.}
    \label{fig:pmt_location}
  \end{subfigure}
  \hfill
  \begin{subfigure}[t]{0.48\linewidth}
    \centering
    \includegraphics[width=\linewidth]{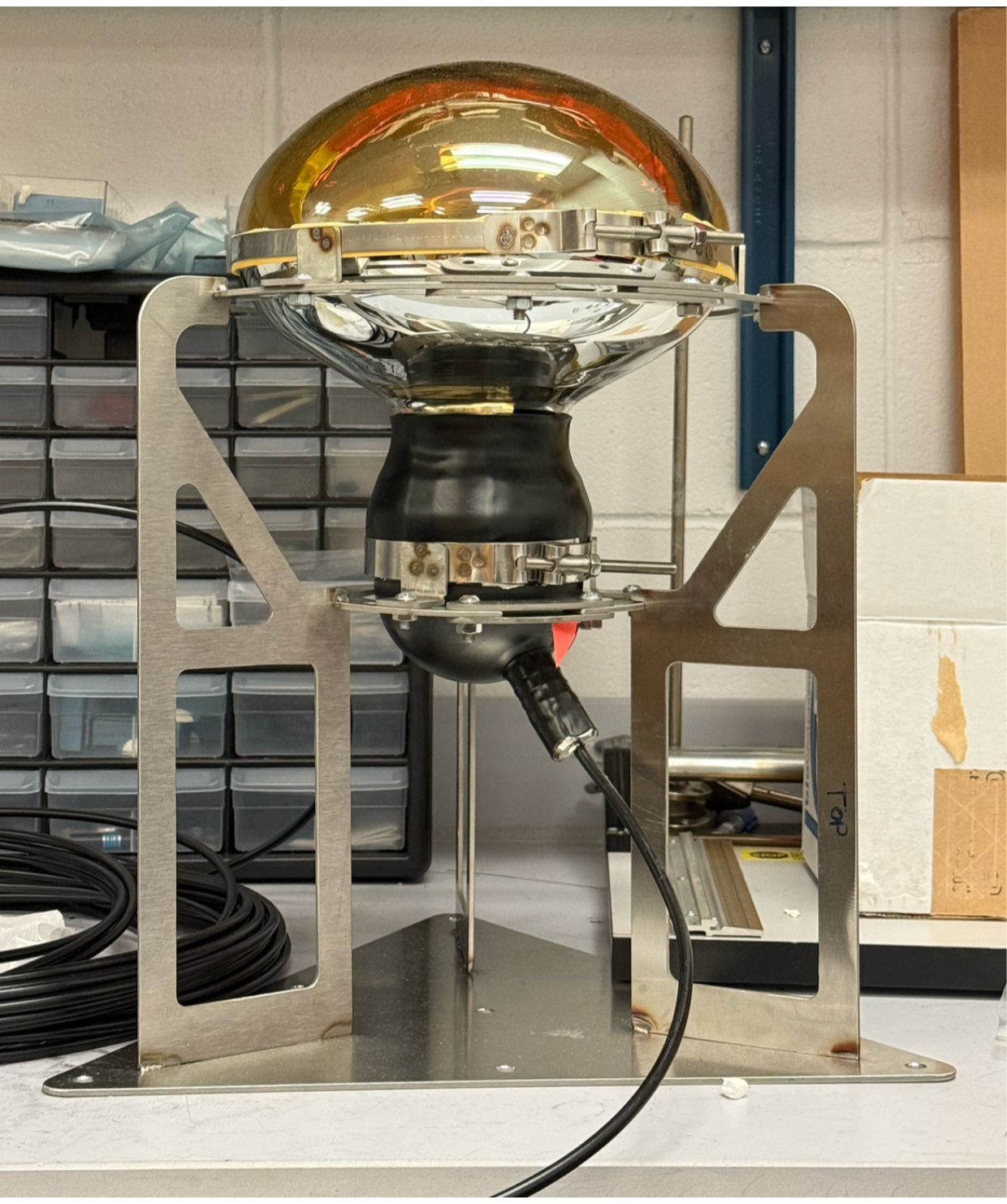}
    \caption{The PMT assembly, showing a 10-inch Hamamatsu R16367 PMT mounted within its custom-fabricated stainless steel support frame. The frame clamps around the equator of the PMT to hold it in place.}
    \label{fig:pmt_frame}
  \end{subfigure}
  \caption{Photographs of the PMT arrangement inside the detector and the PMT support structure.}
  \label{fig:pmts}
\end{figure}

\begin{figure}
    \centering
    \includegraphics[width=0.8\linewidth]{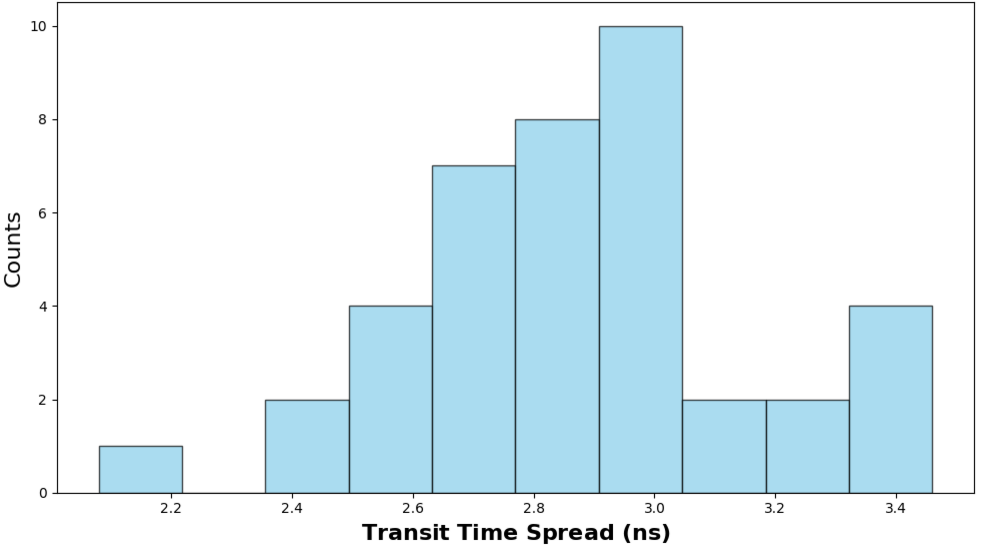}
    \caption{The transit time spread distribution for all the PMTs.}
    \label{fig:tts}
\end{figure}

Before installation, each PMT underwent a detailed characterization of its gain and transit time spread (TTS).  The PMT gain is defined as the amplification factor relating the number of electrons collected at the anode to the initial photoelectron generated at the photocathode. The gain was determined by measuring the single photoelectron (SPE) charge spectrum. For this measurement, each PMT was placed inside a light-tight dark box and illuminated using a pulsed LED system (Thorlabs LED430L, 430~nm) operated at low intensity to ensure predominantly single-photon conditions. This bench-top LED system was used only for the pre-installation tests. 
The PMT output charge was recorded using a Tektronix 6 Series oscilloscope, and the gain was extracted from the mean position of the SPE peak in the measured charge spectrum. Based on these measurements, the nominal high voltage (HV) for each PMT was selected to normalize the gain to approximately $10^{7}$.
The TTS was measured using a $^{90}$Sr beta source coupled to a small acrylic plate to produce a Cherenkov signal. A fast reference PMT (Hamamatsu R1635, 10~mm diameter) with sub-nanosecond timing resolution was mounted on top of the plate to provide a prompt timing reference. This is the same model as the calibration-source PMT described in Sec.~\ref{sec:calibration}, but a separate unit. The time difference between the signals from the R16367 PMT under test and the reference PMT was recorded, and the resulting distribution was fitted with a Gaussian function. 
The Full-Width at Half-Max (FWHM) was taken as the TTS for each PMT.
The measured average TTS was $2.8~\mathrm{ns}$, with a variation of less than $1.0~\mathrm{ns}$ across the tested PMTs. Fig.~\ref{fig:tts} shows the TTS distribution across all the PMTs. 
This is consistent with the manufacturer's typical specification of 3.4 ns (FWHM) for the R16367.
A continuous, long-term calibration of the PMTs is described in Sec.~\ref{sec:calibration}. In addition, detailed studies of the PMT response uniformity were performed. The LED source was positioned at various locations and incident angles on the photocathode surface to evaluate the dependence of gain and timing performance on photon position and angle of incidence. Since external magnetic fields can affect PMT performance, these measurements were conducted inside a custom-designed Helmholtz coil used to cancel the Earth's magnetic field. Three PMTs were selected as a representative test sample and placed at the center of the coil during the measurements. The results indicate that the TTS remained stable, and the gain variation was within 10\% across different illumination positions and incident angles. 

The scintillation time profile of WbLS has been measured in previous work. At the percent-level loadings used here, the emission is dominated by a fast component with a decay time of about 2~ns and a sub-nanosecond rise time~\cite{D0MA00055H,Caravaca2020}. The average TTS of our PMTs, 2.8~ns (FWHM), is comparable to this decay time. Event-by-event separation of Cherenkov and scintillation photons by timing alone is therefore not a goal of this detector. The PMTs were instead chosen for their large area, WbLS compatibility, and cost, which are the relevant factors for a scalability demonstrator. The detector relies on the geometric in-ring/out-ring separation described above. Timing-based separation with fast photosensors is being pursued in complementary demonstrators~\cite{EoS}. A dedicated timing analysis of the 30-ton data is deferred to a future publication.

\begin{figure}
    \centering
    \includegraphics[width=0.6\linewidth]{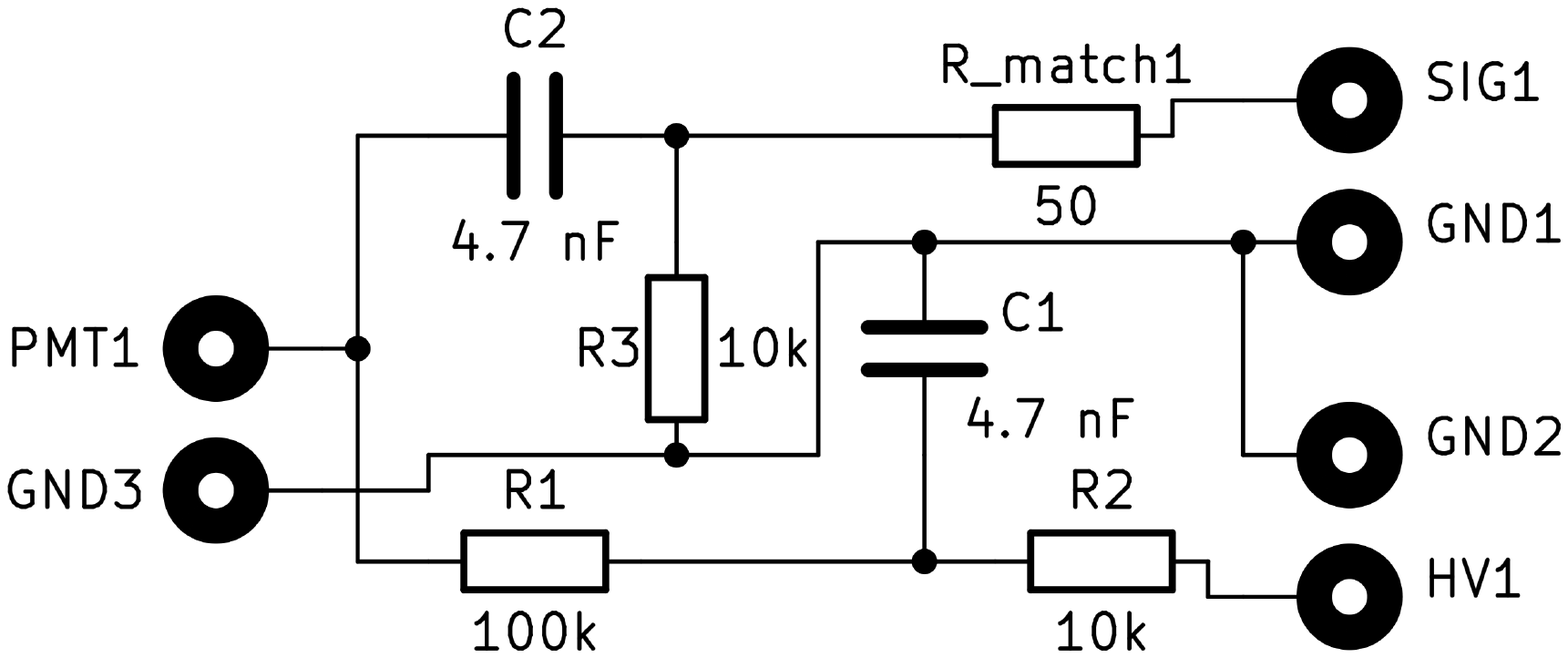}
    \caption{Schematic diagram of the PMT splitter circuit. 
    The circuit decouples the readout signal from the PMT input (PMT1) from the high voltage (HV1). 
    The signal is AC-coupled through a 4.7 nF capacitor (C1) to the signal output (SIG1), which is matched with a 50 $\Omega$ resistor (R\_match1). 
    The HV path is formed by a resistive network, including a 100 k$\Omega$ resistor (R1) and two 10 k$\Omega$ resistors (R2, R3).}
    \label{fig:splitter_box}
\end{figure}

A custom stainless-steel mechanical support frame was designed and fabricated to mount each PMT. Figure~\ref{fig:pmt_frame} shows the assembled PMT–frame unit. The completed PMT assemblies were subsequently installed on the internal support structure of the tank following the geometry described earlier, as illustrated in Fig.~\ref{fig:pmt_location}. Each PMT is read out using a single shielded high-voltage (SHV) cable that supplies the operating high voltage and simultaneously transmits the PMT output signal. A custom-designed splitter box is used to decouple the signal from the high-voltage line; a schematic of the splitter circuit is shown in Fig.~\ref{fig:splitter_box}. The splitter circuit is housed in a 3D-printed plastic enclosure, which is additionally wrapped with an aluminum foil layer to provide electromagnetic shielding from external noise.

Since the PMTs are submerged in WbLS, the integrity of the PMT seals and the stability of the PMT response during exposure to the liquid were important considerations. A two-month soaking test was therefore conducted using a representative PMT continuously submerged in WbLS. During this period, its impedance, gain, and transit time spread (TTS) were monitored. No change in impedance was observed, and no statistically significant variation in the monitored PMT performance parameters was detected. These results demonstrate the integrity of the PMT seals and the stability of the measured PMT response during the two-month exposure to WbLS.

\subsection{Data Acquisition} 
\label{sec:daq}

The 30T DAQ system is similar to that of the 1T prototype. All PMT signals are decoupled from HV signals through custom splitter boxes before being digitized by CAEN V1730S digitizers\footnote{\blue{CAEN S.p.A., Viareggio, Italy, \url{https://www.caen.it}. The V1730S and V1740 digitizers, the V2718 VME bridge, the A3818 controller card, the A7236 HV boards, and the SY5527 mainframe are all supplied by CAEN.}}. The V1730S boards use 14-bit ADCs with a sampling rate of 500 MSPS and each has 16 channels. We have three V1730S digitizers in our DAQ system. Additionally, signals from the PMTs, Hamamatsu R1847 devices optically coupled to the scintillator paddles for muon tagging, are digitized using a CAEN V1740 digitizer. The V1740 has 64 channels, uses 12-bit ADCs, and has a sampling rate of 62.5 MSPS. The clock from the master V1730S is propagated to the other three boards through a daisy chain. Clock phase shifts between all boards were carefully synchronized up to 400 ps. Communication and data flow between the digitizers and the DAQ computer is done via a CAEN V2718 VME bridge board connected by an optical link to a CAEN A3818 PCIe controller card installed in the DAQ computer.

All PMTs and scintillating paddles are powered by CAEN A7236 HV supply boards, connected to the splitter boxes via SHV cables. The supply boards are in a CAEN SY5527 HV mainframe, which is connected to the DAQ computer by Ethernet, allowing for communication to the supply boards. The HV supplies are programmed using the CAENGECO2020 software, which also performs logging of channel voltages and currents for monitoring purposes. 

\begin{figure}
    \centering
    \includegraphics[width=1\textwidth]
  {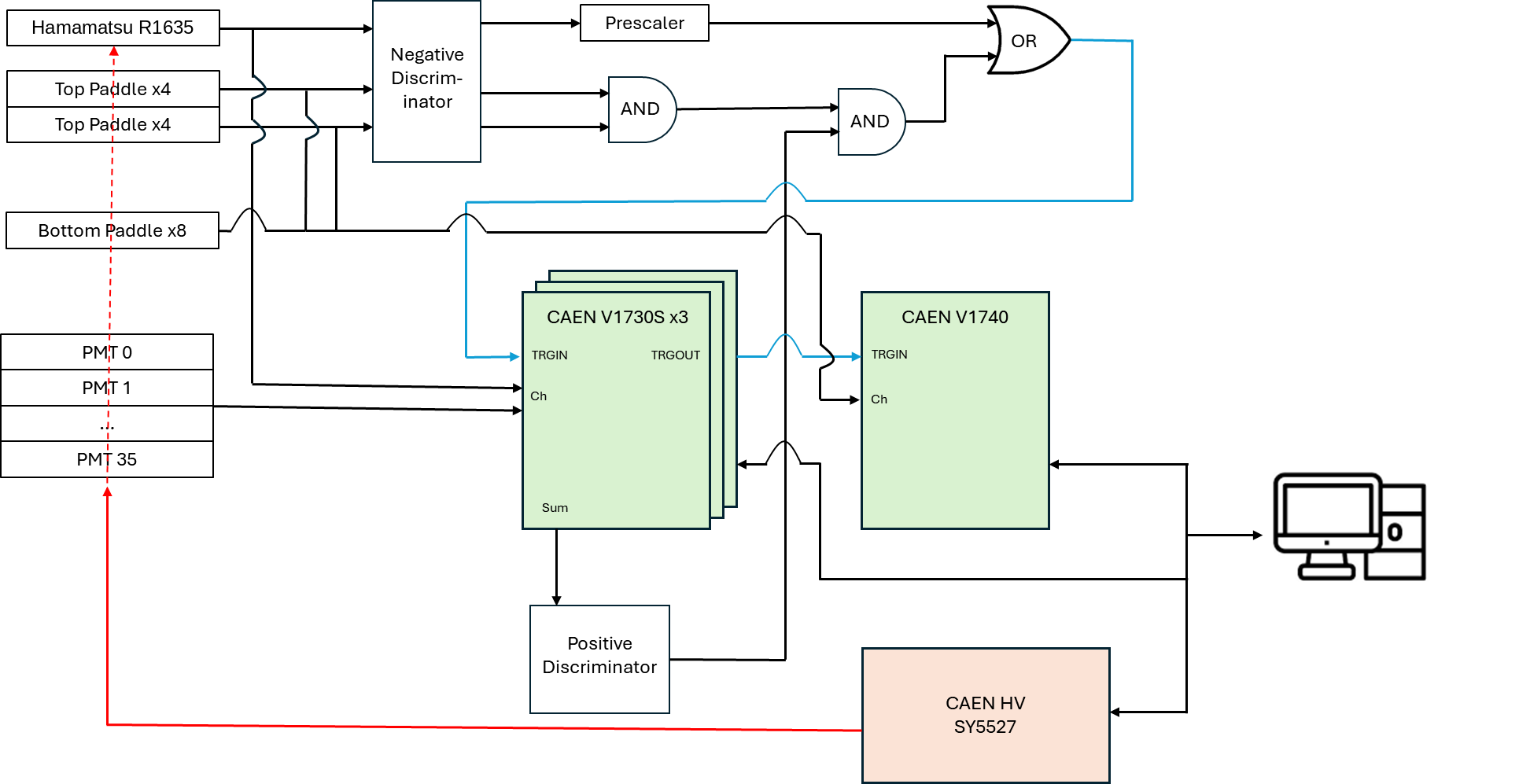}
    \caption{A diagram of the readout system for the 30T WbLS demonstrator. A trigger is formed by requiring a coincidence of two layers of scintillating paddles above the detector, along with a signal from the majority trigger. The majority trigger sends a signal when at least 5 bottom PMTs cross a specified voltage threshold. In addition to this requirement, a trigger may be caused by the Hamamatsu R1635 from the $^{210}\mathrm{Pb}$ calibration source setup. The resulting trigger signal is sent to the first digitizer external trigger input and is propagated via daisy chain to the other boards. All PMT signals are digitized on CAEN V1730S digitizers, while the paddle signals are digitized by a CAEN V1740 digitizer. Finally, high voltage to the PMTs and paddles is provided by CAEN HV supplies in a CAEN SY5527 mainframe. The mainframe and digitizers are connected to the DAQ computer for data readout, monitoring, and configuration. }
    \label{fig:daq}
\end{figure}

The 30T detector employs multiple trigger configurations depending on the measurement being performed, including triggers based on scintillating paddles, PMT majority signals, and an independent $^{210}$Pb calibration source. For cosmic muon measurements, eight scintillating paddles are arranged above the detector in two layers of four adjacent paddles. When a muon passes through both layers, a coincidence signal is formed. A trigger is then generated when this coincidence occurs simultaneously with a PMT majority trigger. The PMT majority trigger is produced by applying thresholds to all bottom PMT channels in the digitizer.
A signal related to the number of channels that cross their thresholds is output by the digitizer. This signal is then discriminated against a threshold set such that at least 5 bottom PMTs must have recorded a hit in an event before the majority signal is generated. In addition to the AND of the coincidence of top paddles and the majority trigger signal, a trigger signal is generated if the $^{210}\mathrm{Pb}$ calibration source setup is triggered. The trigger signal is input into the first V1730, which propagates it to the other digitizers in a daisy chain. The delay caused by the daisy chain is stable and is subtracted in offline analysis. Finally, to tag muons which pass through both the top and bottom of the detector, as opposed to muons which exit through the side of the tank, another eight scintillating paddles are mounted beneath the detector. Fig.~\ref{fig:daq} shows the overview of the trigger and DAQ system.

A DAQ software based on the ToolDAQ framework was developed for use in the 1T and 30T demonstrators. ToolDAQ, written in C++, is lightweight, modular, and highly customizable~\cite{Richards:2019lkb}. The 30T DAQ software utilizes CAEN C++ libraries to program and read data from the digitizers. The data is written out to binary files before being processed offline and written to ROOT files.

\section{Detector Circulation System} 
\label{sec:circulation}
\begin{figure}[htbp]
    \centering
    \includegraphics[width=1.0\textwidth]{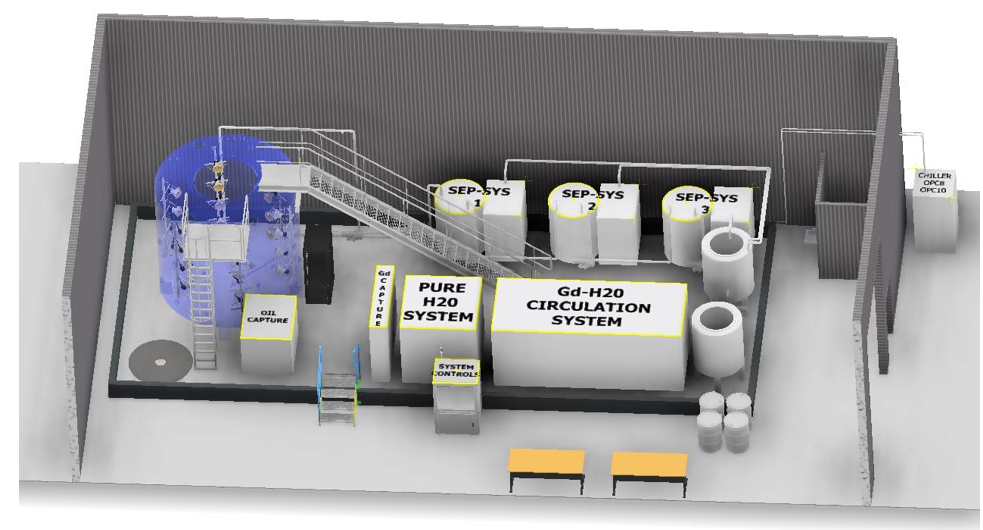}
    \caption{Overview of the 30-ton detector facility layout. The main detector tank is shown in blue on the left. The adjacent systems include the Pure H2O and Gd-H2O circulation skids, three separation/purification systems (SEP-SYS 1, 2, 3, nanofiltration), and capture units for gadolinium and oil.}
    \label{fig:facility_layout}
\end{figure}

\begin{figure}[htbp]
    \centering
    \includegraphics[width=0.5\textwidth]{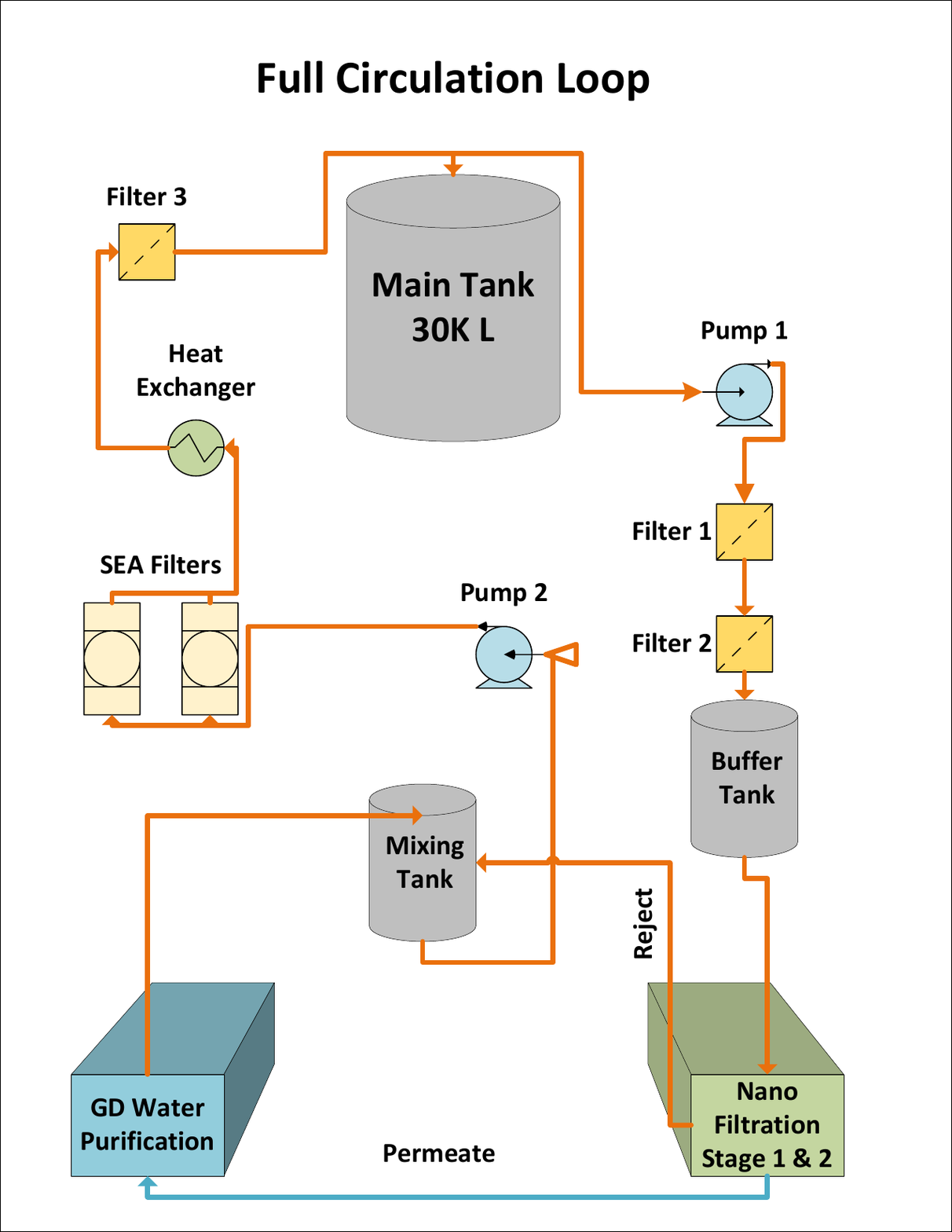}
    \caption{Schematic layout of the purification system.}
    \label{fig:full_cir}
\end{figure}

The 30-ton detector is supported by a sophisticated liquid handling and purification plant housed within a dedicated facility, as depicted in Figures~\ref{fig:facility_layout} and~\ref{fig:full_cir}. The entire detector and circulation systems are in a secondary containment system for safety. The integrated system is designed to test the long-term stability and optical clarity of the WbLS medium under controlled purity conditions. 
Continuous circulation is a departure from the essentially static operation of conventional liquid scintillator detectors such as Borexino~\cite{BOREXINO:2023ygs} and KamLAND~\cite{PhysRevLett.90.021802}. It carries potential risks, including mechanical disruption of the micelles, radon ingress, and thermal gradients. No evidence of micelle degradation was seen during more than two years of continuous circulation in the 1-ton prototype~\cite{1ton-paper}. Characterizing such effects at a larger scale is a primary goal of this demonstrator. The preliminary stability results in Secs.~\ref{sec:calibration} and~\ref{sec:performance} show no degradation of the optical response during the initial months of operation with circulation. 
The primary purification loop is centered on a series of multi-stage purification modules, including an NF system, a Gd purification system, and a resin-based sequential ion-exchange array (SEA). The system is designed with flexible flow paths, allowing different circulation loops to bypass or route through selected components as needed, as illustrated in Figure~\ref{fig:full_cir}. During commissioning, the NF and Gd purification systems were not implemented.


Together, these interconnected subsystems
form a comprehensive processing plant, allowing the detector medium to be continuously circulated, purified, and chemically tuned, demonstrating the robust and scalable technologies required for future kiloton-scale WbLS detectors.


The circulation system is constructed from a combination of  316 Stainless Steel and Polypropylene. Connections are made using a mixture of flanged, sanitary Tri-clamp, NPT, and plastic fusion welds.
Circulation commences with a 3-inch stainless steel pipe, submerged approximately 4 feet below the water level, which draws fluid from the top of the tank. This pipe then transitions to polypropylene and directs the fluid through a Finish Thompson SP22 Pump\footnote{\blue{Finish Thompson Inc., Erie, PA, USA, \url{https://www.finishthompson.com}}}.
In the primary circulation loop, fluid is propelled through a series of fine micron filter housings\footnote{0.1 micron pore size filter membranes are used}, resin bead housings, and a heat exchanger\footnote{The heat-exchange system, working together with the chiller, must maintain the liquid scintillator at $16^\circ\mathrm{C}$ under normal and expected operating conditions.
} before re-entering the tank via a 1.5-inch stainless steel pipe, submerged approximately 11 inches below the water level.

An alternative circulation loop allows the initial SP22 pump to push fluid into a series of fine micron filter housings before it enters the first 200-gallon subtank, referred to as the "mixing tank." From the mixing tank, a second pump draws fluid from the drain and propels it through the resin bead housings, heat exchanger, and a fine micron filter housing before it returns to the main tank. This circulation loop is essential for chemical injections, as it facilitates the injection of external fluids without introducing light leaks into the main tank and enables the pre-mixing of chemicals prior to their introduction into the main tank.


The complete circulation loop integrates the nanofiltration system, the gadolinium purification system, and the sequential exchange array with the central detector vessel. An initial pump drives the fluid through a series of fine micron filter housings and into a feed tank for the NF system. The reject stream from the NF system is returned to the mixing tank, while the permeate is routed to the gadolinium purification system. The output of the gadolinium system is likewise returned to the mixing tank. Fluid from the mixing tank is then pumped through the resin-based SEA, followed by a heat exchanger and a final fine micron filter housing, before being returned to the main detector vessel.

The main tank also features a drain line, which can be utilized to draw fluid from the bottom of the tank for circulation or to provide feed fluid. A dedicated pump is available for system draining.
A pure water system is accessible within the facility to provide Omega-18~\footnote{Omega-18 water is ultrapure water with a resistivity near 18 M$\Omega$cm, close to the theoretical limit for pure water at room temperature.} highly pure water as needed. The system's output can be directed to the input line of the main tank, and it includes an input for recirculation, allowing for the maintenance of water quality. This recirculation receives its feed from the main tank's drain line. The resin-based SEA is currently commissioned, whereas the NF and Gd systems are ready and will be integrated into the circulation system during future runs.

\subsection{Nanofiltration System}

There is compelling evidence that stainless steel exposed to gadolinium chloride will leach contaminants into the water that affect optical transparency in the ultraviolet and optical region of the spectrum~\cite{coleman}, likely from dissolved iron ions. Thus, a system is needed that would allow removal of such ions without disrupting the micelles or affecting gadolinium concentration. The strategy adopted has two major steps: (1) separate the micelles from the rest of the WbLS using NF with a large pore membrane, and (2) remove the remaining free organics using a tighter NF membrane in order to pass the remaining liquid through a system that can remove iron and other optical absorbing ions without removing the gadolinium. Note that this is best done in two steps, as using a single tight filter stage would likely lead to significant fouling due to the large size of the micelles (several nanometers) compared to the typical dissolved ion (sub-nanometer scale).

\begin{figure}[htbp]
    \centering
    \includegraphics[width=1.0\textwidth]{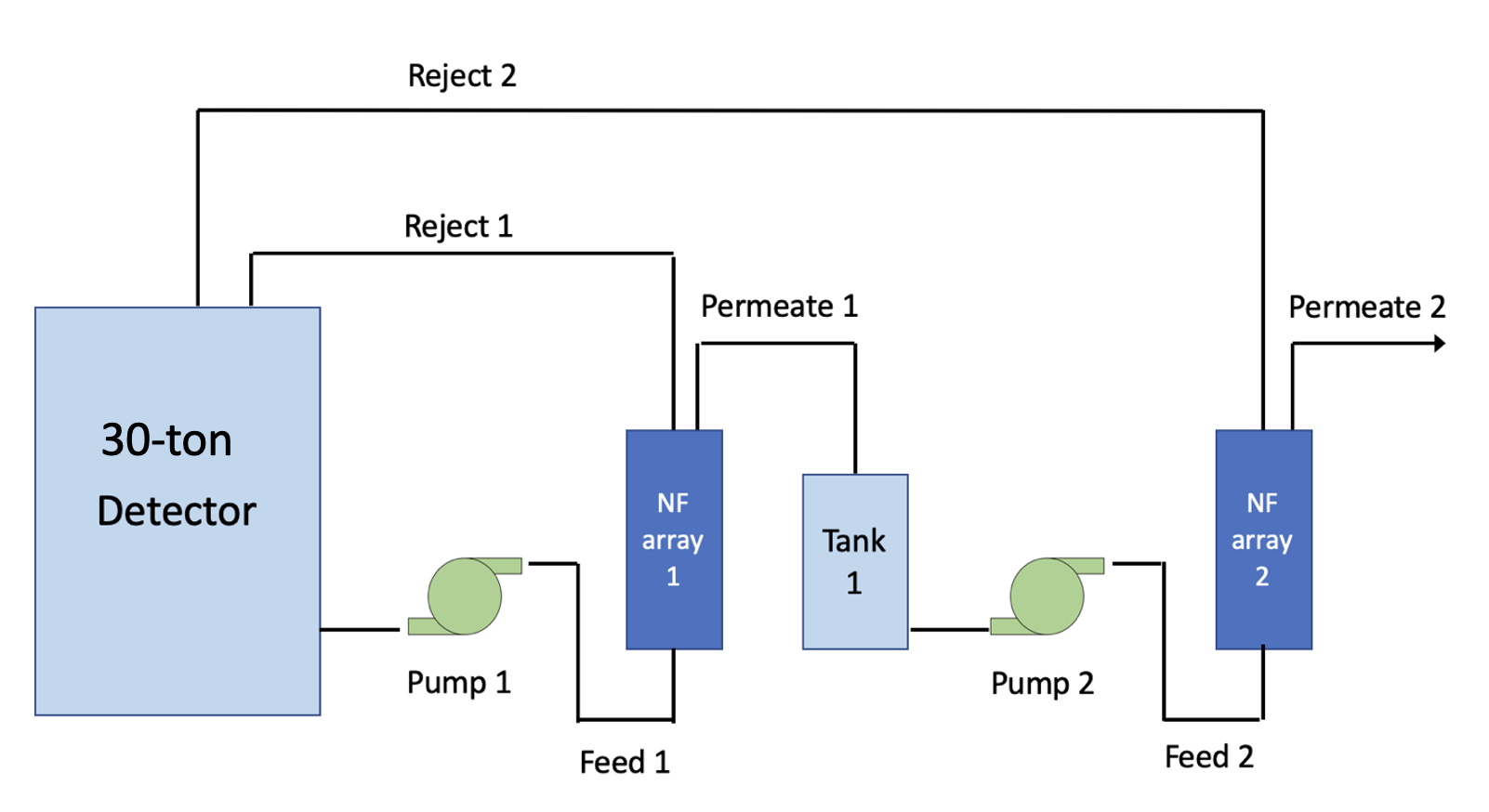}
    \caption{Conceptual layout of the Nanofiltration System}
    \label{fig:NFsystem}
\end{figure}

Figure~\ref{fig:NFsystem} depicts a conceptual flow diagram for the organics removal system. Pump 1 takes WbLS from the detector tank and sends it through a first-stage NF array to remove micelles. Reject 1 from this stage, containing the removed micelles, is returned to the detector, while Permeate 1, containing free organics and optical contaminants, is sent to an intermediate tank. Pump 2 pressurizes this liquid to send it through a second NF stage that removes most of the remaining organics and returns them to the detector as Reject 2. Permeate 2 then goes to a system that can remove optical contaminants, such as a Molecular Band-Pass filter system or a SEA as described below.

\begin{figure}[htbp]
    \centering
    \includegraphics[width=1.0\textwidth]{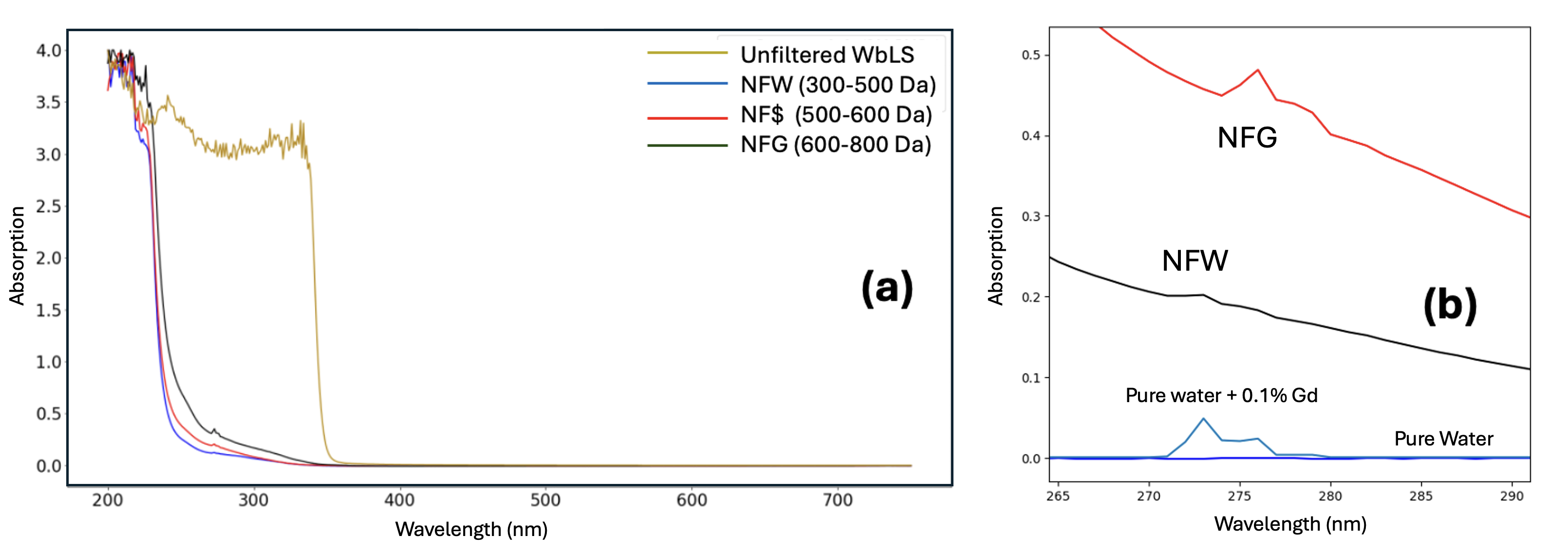}
       \caption{(a) Absorption spectra in a 10-cm cell for unfiltered Gd-loaded WbLS and for the permeate from three candidate Stage 1 commercial nanofilters with increasing MWCO: NFW (300-500 Da), NF\$ (500-600 Da), and NFG (600-800 Da). (b) Absorption spectra for the permeate from NFG (red) and NFW (black) in the region of the gadolinium absorption peaks. }
    \label{fig:Stage1}
\end{figure}

\subsubsection{Identification of NF Candidates}

A key part of the R\&D for the system design was to identify appropriate NF membranes for both stages. Commercial membranes come in many pore sizes, usually characterized by the Molecular Weight Cut-Off (MWCO). The MWCO is the molecular weight in Daltons (Da) of the largest molecule that would be able to pass through the membrane. The MWCO is often given as a range due to the fact that many manufacturing processes do not allow strict control of the pore size, but rather result in a distribution of sizes.

Benchtop tests conducted at the University of California, Davis, using a MaxiMem single-stage NF system from Prozesstechnik GmbH, were used to identify a first-stage filter that would remove micelles but not gadolinium sulfate or iron contaminants and that was also compatible with WbLS. Figure~\ref{fig:Stage1}(a) shows the 10-cm cell relative optical absorption spectra compared to pure water for unfiltered WbLS plus permeate solutions from three candidate NFs of differing MWCO. The unfiltered spectrum is characterized by a sharp absorption edge at 340 nm due to the PPO inside the micelles of the WbLS. The apparent absorption plateau around 3.0 is due to instrument saturation and not physical.~\footnote{The actual absorption is larger, but the instrument sensitivity is only 3.0-3.5 depending on the wavelength. Thus the instrument simply displays all larger numbers as being at or near the maximum sensitivity.} The other three curves are from a single pass through three commercial candidate filters from Synder Filtration, arranged in order of smaller MWCO. To put these in context, the coordinated MW for $Gd^{+3}$ is in the range of 301-320 Da, while $Fe^{+3}$ is about 164 Da.

The NFG (MWCO 600-800 Da) spectrum shows a very significant reduction on the micelle encapsulated PPO, indicating that almost all of the micelles have been removed. Though difficult to measure, we estimate based on the spectrum that more than 99\% of the micelles have been removed. NF\$ (MWCO 500-600 Da)\footnote{This membrane was custom-fabricated for this study and is not currently available commercially.} and NFW (MWCO 300-500 Da) show slightly more removal, but not significantly more.
This would be expected from the relatively large size of micelles compared to even the NFG MWCO. Figure~\ref{fig:Stage1}(b) shows the spectra for NFG and NFW in the region of the double absorption peaks from the gadolinium ion in the WbLS. Also shown is the spectrum from a water solution of gadolinium sulfate that has 0.1\% by weight of $Gd^{+3}$. By comparing these with the spectra of the NFW and NFG permeate, it can be determined that NFW passes only 17\% of the gadolinium while NFG passes more than 98\%. While in principle it would be acceptable to remove gadolinium in Stage 1 since it will be returned to the detector in Reject 1, the fact that gadolinium passes through the NFG is a good indication that the iron ions will also pass through, which is critical to the process. In addition, it is desirable to use the largest MWCO that effectively removes micelles in order to mitigate fouling in Stage 1. Thus, the NFG membrane was selected for Stage 1.

Using liquid chromatography and molecular mass spectrometry on the NFG permeate, it was determined that the remaining ``free'' organics (organics not encapsulated in micelles) have a MW in the range of 300-600 Da. This was consistent with the fact that the MWCO range for NFG (600-800 Da) would pass these molecules through. Thus, Stage 2 must have a MWCO that would allow filtering these organics out without retaining the iron corrosion contaminants. 

For identifying potential Stage 2 membranes, a Sterlitech CF042 test device with an acrylic flat sheet membrane cell was used. The feed liquid was not WbLS but pure water exposed to 316 stainless steel plates for over two years.
Several potential filters were identified, including the TriSep-40 (TS-40) membrane with MWCO 200 Da, which should effectively remove the remaining free organics but also pass dissolved iron ions. This was tested by passing pure water exposed for more than two years through plates of 316 stainless steel as a feed liquid.

\begin{figure}[htbp]
    \centering
    \includegraphics[width=1.0\textwidth]{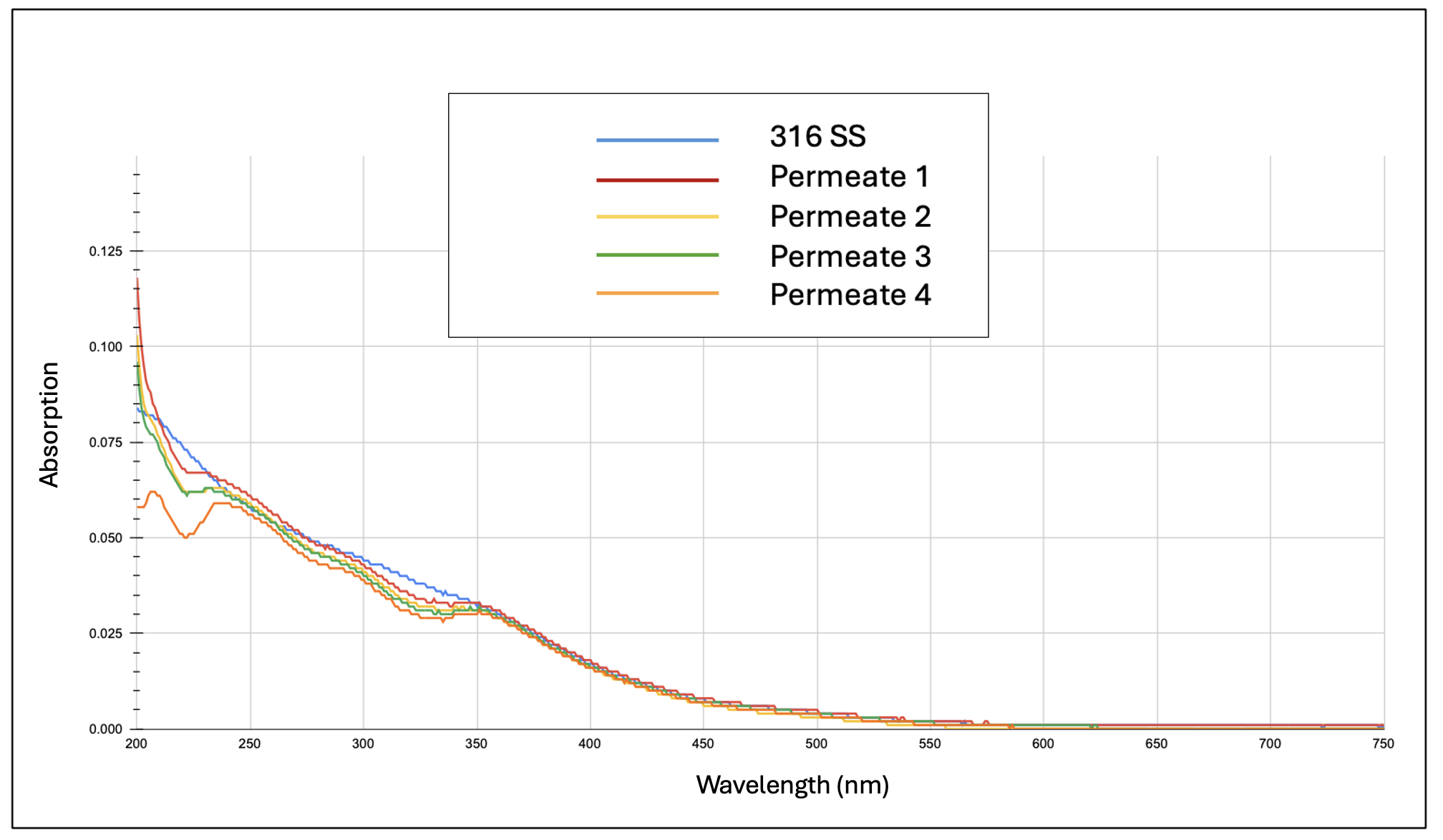}
    \caption{Absorption spectra in a 10-cm cell for water soaked for two years in 316 SS plates passed multiple times through a TriSep-40 filter (MWCO 200 Da). The blue curve is the initial water showing the effect of steel corrosion products on transparency. The other curves are only slightly different, showing that the filter is passing most of the optical contaminants. The origin of the feature below 240 nm is not known, but is well below photocathode sensitivity.}
    \label{fig:Stage2}
\end{figure}

Figure~\ref{fig:Stage2} shows the 10-cm cell relative absorbance of the ``rust water''~\footnote{The water was actually clear to the eye, but this was a convenient name for the liquid.} compared to four successive passes of the liquid through the TS-40 membrane. It can be seen that TS-40 does not significantly retain the optical contaminants from the rust water and thus is a candidate filter to be tested in our NF system for the 30-ton detector, as described below.

\subsubsection{The 30-ton NF System}

It was decided that a very short turnover time~\footnote{The turnover time is defined as the tank volume divided by the system throughput rate} would be prudent in order to enable a broad spectrum of possible operating conditions. For example, a turnover rate of 1 day would require a Permeate 2 flow rate of:
\begin{equation}
\frac{\left( 30 \: m^3 \right) \left( 1000\; L/m^3\right) }{\left( 1\; day\times 1440\; min/day  \right)}= 20.8 \; L/min
\end{equation}

The system specifications for achieving a given turnaround time, the NF system was based on scaling up performance results from the MaxiMem system with a spiral-wound 1812 NFG filter, which has an area of $0.37 \; m^2$. Table~\ref{maximem1} shows the permeate results for running the MaxiMem at roughly a 20 bar pressure across the membrane and varying feed flow rates. The 20 bar feed pressure was selected as it is well below the MaxiMem overpressure safety trip at 40 bars and the NFG 1812 rated maximum cross-membrane pressure of 40.8 bars. The maximum feed flow of 2.40 $L/min$ was based on the typical operating feed rate GFD~\footnote{gallons per square foot per day} of 55-60, which corresponds to $2.1-2.3\; L/min$.

\begin{center}
\begin{table}[h]
\begin{tabular}{c|c|c|c|c}
temperature ($C$) & pressure ($bar$) & feed flow ($L/min$) & permeate flow (L/min) & recovery ratio \\
\hline
20.9 & 20.2 & 2.40 & 0.33 & 0.14 \\
21.1 & 20.7 & 2.00 & 0.32 & 0.16 \\
21.3 & 20.6 & 1.50 & 0.31 & 0.21 \\
21.5 & 20.5 & 1.00 & 0.29 & 0.29 \\
21.5 & 20.5 & 0.50 & 0.29 & 0.58 \\ \hline
\end{tabular}
\caption{Performance of the 1812 spiral-wound NFG filter and the MaxiMem test bench device}
\label{maximem1}
\end{table}
\end{center}

It can be seen that the permeate flow across the membrane at fixed pressure is relatively independent of feed flow, as expected. The 0.58 recovery ratio~\footnote{The recovery ratio is defined as the ratio between the permeate and feed flow rates}  was the maximum one tested, based on vendor information on recovery ratios with solutions of water and lactose, which is similar to our application. This ratio may be varied as experience is gained with the 30-ton system in order to optimize performance.

These measurements were scaled up to a required throughput of 20.8 $L/min$ by using a constant recovery ratio and making the reasonable assumption that flow rates scale by the area of the NF. It was also assumed that Stage 2 performance would be similar to Stage 1. In addition, spiral filters are manufactured in standard sizes which limits the number of practical options.  Thus, a system comprising two parallel 8040 spiral-wound NFG filters (area of $70.6\; m^2$) was selected for Stage 1 and a single 8040~\footnote{Filter form factors are given as xxyy where xx is the diameter in tenths of an inch and yy is the length in inches.} TS-40 spiral-wound filter (area $35.3\; m^2$) for Stage 2. With this combination, the Permeate 1 and Permeate 2 flow rates per square meter would be 0.51 and 0.59 $L/m^2/min$ respectively. This is conservatively less than the Stage 1 flow rate of 0.78 $L/m^2/min$ actually achieved in the MaxiMem tests. 
Thus a Stage 2 pump capable of delivering $(20.8\; L/min)/(0.58)= 35.9\; L/min$ (9.5 $gpm$) and a Stage 1 pump capable of delivering $(35.9\; L/min)/(0.58)= 61.9\; L/min$ (16.4 $gpm$) are required.

Due to the need for strict control of the materials so as to be compatible with WbLS, the system was mostly constructed of stainless steel with fiberglass filter housings, which were soak tested to verify they would not degrade optical quality due to leaching. In addition, all O-rings were required to be made of Viton and all pump internals to be stainless steel or plastics known to be compatible. Several vendors were identified who could build such a custom system, and Synder Filtration was selected for the engineering design and fabrication.

\begin{figure}
\centering
\includegraphics[width=1.0\textwidth]{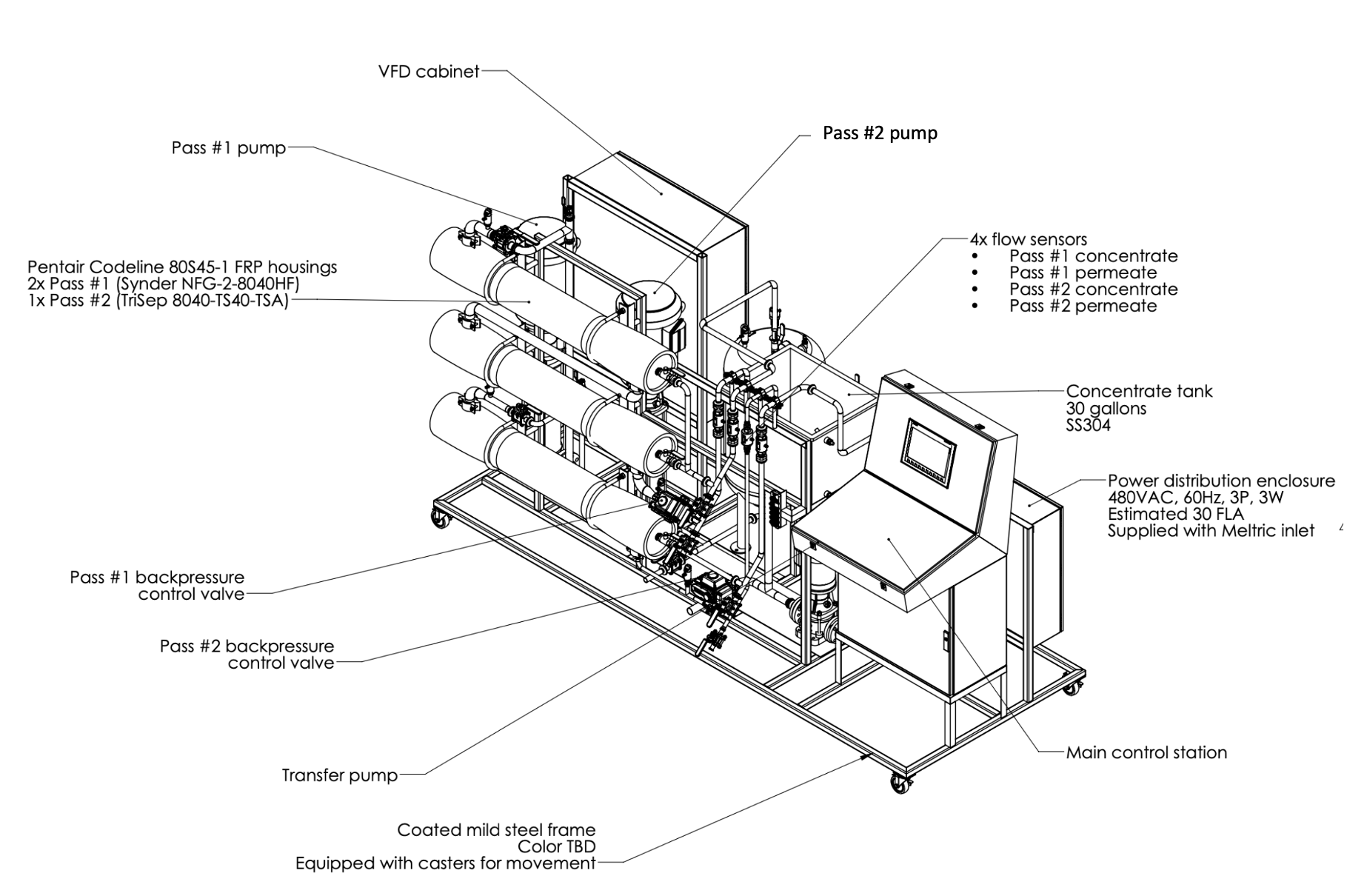}
\caption{Isometric view of the NF system.}
\label{fig:NFisometric}
\end{figure}

Figure~\ref{fig:NFisometric} depicts an isometric view of the 30-ton NF system as built with two Stage 1 8040 NFG filters and one Stage 2 8040 TS-40 filter. The system is fully automated, with a programmable logic controller (PLC)-based control system that is integrated with the overarching 30-ton control system. An intermediate 370 $L$ tank collects Permeate 1 for re-pressurization to Stage 2. A nitrogen blanket system is provided to allow oxygen-free operation if needed. The Stage 2 permeate is directed to an input tank to the Gd-H$_2$O system described in the next section.

\subsection{Gd-H$_2$O system}

After the WbLS has been largely removed from the recirculating fluid stream by the nanofiltration system, the next step is to remove all remaining impurities in the water \emph{except} for the dissolved gadolinium sulfate itself.  In order to accomplish this, a truly selective filtration system tuned for gadolinium sulfate is needed.  The heart of this "Gd-H$_2$O" system is known as a "Molecular Band-pass Filter".  

\begin{figure}[htbp]
    \centering
    \includegraphics[width=1.0\textwidth]{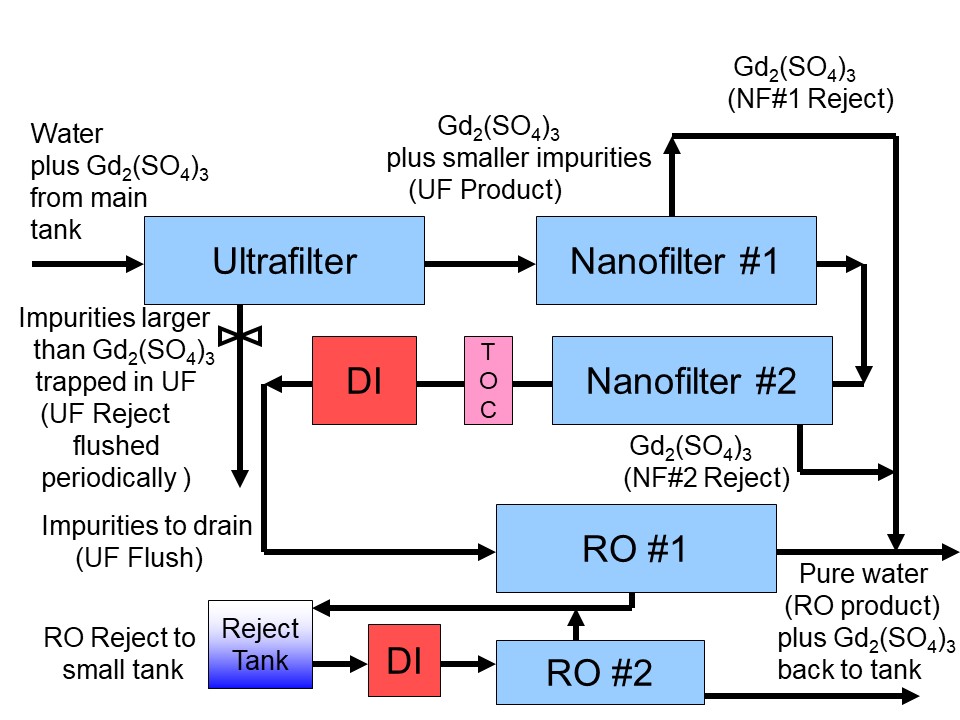}
    \caption{Schematic layout of BNL's Molecular Band-pass Filter system. In order of appearance in the flow, the main components are as follows: an Ultrafilter rejects and traps all impurities larger than aqueous Gd$^{3+}$ and (SO$_4$)$^{2-}$ ions; Nanofilters \#1 and \#2 are tuned to specifically reject aqueous Gd$^{3+}$ and (SO$_4$)$^{2-}$ ions and return them to the 30-ton tank while passing smaller impurities; TOC (Total Organic Carbon) uses UV light to ionize compounds containing carbon atoms; DI (deionization) removes the remaining ions from solution, including carbon compounds ionized by the TOC lamp; finally, RO \#1 and RO \#2 (reverse osmosis) reject and trap any remaining impurities smaller than aqueous Gd$^{3+}$ and (SO$_4$)$^{2-}$ ions, returning the RO product -- ultrapure water -- to the 30-ton tank.}
    \label{fig:band-pass}
\end{figure}

Invented and prototyped nearly two decades ago at the University of California, Irvine, a series of staged filter membranes returns aqueous Gd$^{3+}$ and (SO$_4$)$^{2-}$ ions to the circulation stream while trapping both larger and smaller impurities. This is shown schematically in Figure~\ref{fig:band-pass}.

This fluid-based system is entirely analogous to a classic electronic band-pass circuit, where only frequencies within a certain range are accepted, as utilized in old-fashioned radio or TV tuners.  Here, instead of allowed frequencies, the staged filter membranes -- ultrafilters (UF), nanofilters, and reverse osmosis filters (RO) -- restrict the retained ions to a narrow size band, which in this case has been tuned to Gd$^{3+}$ and (SO$_4$)$^{2-}$.  The reject of the primary RO is recycled through a secondary RO loop to avoid loss of working fluid.

The BNL band-pass is the fourth such operational system in the world, following the UCI prototype (2009), the EGADS 200-ton  gadolinium testbed in Japan~\cite{egads} (2010), and the XENONnT dark matter experiment's veto shield~\cite{xenonnt} in Italy (2023).

\subsection{Sequential Exchange Array (SEA)}
\label{subsec:sea}

The 30-ton testbed tank is constructed from 316L stainless steel, an alloy composed primarily of iron (70\%), chromium (18\%), and nickel (8\%). During extended exposure to water, these metallic constituents can leach into the surrounding medium as dissolved ionic species---most commonly Fe$^{2+}$/Fe$^{3+}$, Cr$^{3+}$, and Ni$^{2+}$---particularly if the passive oxide layer on the steel surface becomes degraded or non-uniform.

To enhance corrosion resistance and minimize surface contamination, the tank underwent a pickling treatment using a 30\% Citranox--water solution. This process removes native oxide layers and promotes the formation of a chromium-rich passive film, improving resistance to further corrosion. Nevertheless, long-term water contact can gradually deteriorate the passive layer, resulting in trace-level metal-ion release.

To mitigate the presence of leached ions, a resin-based filtration unit termed the SEA was developed. The SEA consists of a stainless-steel cylinder packed with a mixture of commercial ion-exchange resins. These resins exhibit porous microstructures, large internal surface areas, and abundant hydroxyl functional groups, enabling efficient binding of metal ions---particularly Fe$^{3+}$---through ligand-exchange and surface-complexation mechanisms.

\subsubsection{Benchtop Iron-Removal Assessment}

Fe-removal performance was assessed using GdWbLS spiked with iron. The Fe-spiked solution was prepared by diluting FeCl$_{3}$ in an acidic aqueous matrix to a concentration of 50~ppm, followed by secondary dilution into GdWbLS to achieve the 10~ppm target level. A 40-g vial was filled with the spiked GdWbLS and amended with a known mass of the resin mixture. The vial was continuously agitated for 24 hours, followed by a 30-minute settling period prior to analysis.

\begin{figure}[htbp]
    \centering
    \includegraphics[
        width=0.80\textwidth,
        trim=50 400 50 50,
        clip
    ]{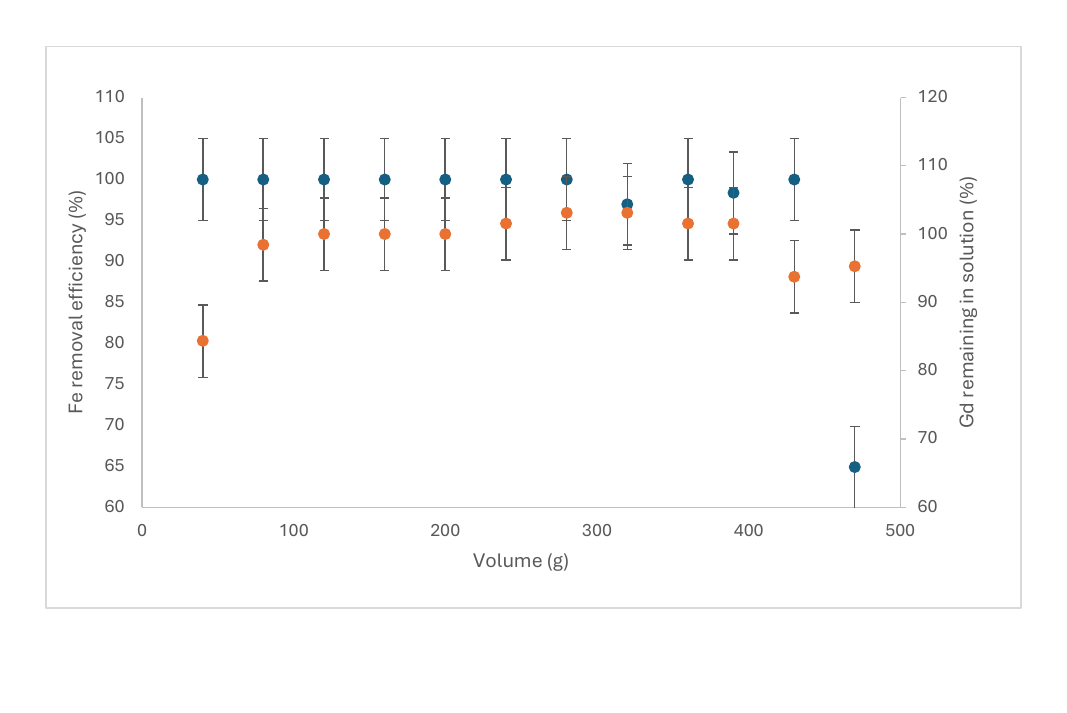}
    \caption{SEA performance using Fe-spiked GdWbLS: iron removal efficiency (blue) and gadolinium remaining percent (orange).}
    \label{fig:SEA_fe}
\end{figure}

Under these conditions, the resin exhibited strong Fe-removal capability without the extraction of organic components. Follow-up tests performed with freshly spiked GdWbLS showed that the same batch of resin could treat up to 430~mL of solution before the Fe-removal efficiency fell below 70\% as shown in Fig.~\ref{fig:SEA_fe}. During the initial treatment, the resin removed approximately 3\% of the Gd content, indicating partial non-selective binding. However, no additional Gd loss was observed in subsequent treatments, suggesting that the resin quickly reached Gd saturation and maintained selectivity for Fe ions thereafter.

Fe detection in solution was performed using a Shimadzu EDX system (energy-dispersive X-ray analysis). Based on these measurements, the total Fe-removal capacity of the resin was determined to be 14.33~mg Fe per gram of mixed resin.

\subsubsection{Scale-Up Projection}

The 30-ton WbLS demonstrator contains approximately 7200~gallons of WbLS. Assuming an Fe-leach concentration of 10~ppb and continuous circulation at 25~gallons per minute (24~hours per day), the total processed volume over a three-month period is:
\[
V_{\rm total} \approx 12.7 \times 10^{6}~\mathrm{L}
\]
This corresponds to an estimated Fe load of:
\[
M_{\rm Fe} \approx 122.7~\mathrm{g}
\]

Using the benchtop-derived removal capacity:
\[
\mathrm{Capacity} = 14.33~\mathrm{mg~Fe/g~resin}
\]
the required resin mass to remove the total Fe load is:
\[
m_{\rm resin} = \frac{122.7~\mathrm{g}}{0.01433~\mathrm{g/g}}
\approx 8.5~\mathrm{kg}
\]

These results demonstrate that the SEA provides an effective and scalable approach for long-term metal-ion control in large GdWbLS detectors. The resin mixture offers substantial Fe-removal capacity while exhibiting minimal interference with Gd concentration after initial saturation. The SEA contains 8.5 kg of this resin mixture, sufficient for a three-month service interval at the current circulation rate. Bulk production keeps the projected 10-year resin cost below \$3,000, resulting in a low lifecycle operating cost. This performance supports stable optical and chemical conditions during continuous operation of large-scale WbLS systems.

\section{Slow Control system} 

\begin{figure}
    \centering
    \includegraphics[width=1\textwidth]
  {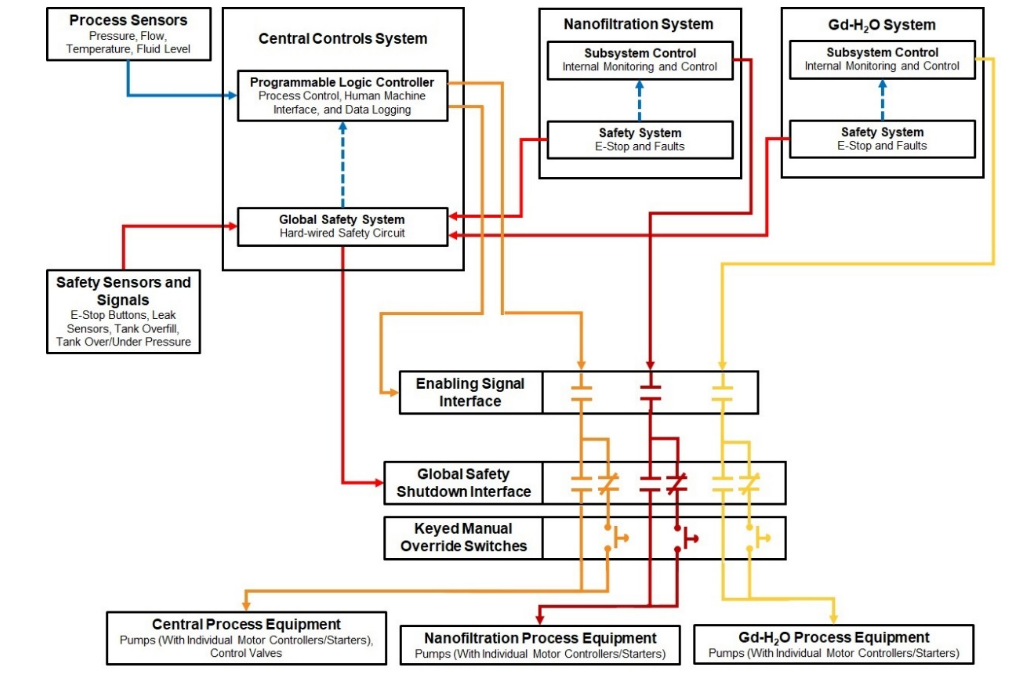}
    \caption{Conceptual diagram of the WbLS 30–ton detector central control system.}
    \label{fig:slow_control}
\end{figure}

The WbLS recirculation system of the 30-ton detector requires continuous monitoring and control of various process parameters to ensure optimal performance and safe operation. The name “slow control system” arises from the fact that the system focuses on parameters that evolve over longer time scales compared to the physics events recorded by the detector DAQ system. The 30-ton detector central control system is two-tier: 1) a process control system maintains stable operation of the detector recirculation system and consists of a PLC to monitor and control all inputs and outputs provided by the WbLS recirculation equipment, an operator interface to initiate either automatic or manual control, and a data logger and visualizer. 2) A global safety control system monitors a set of specific inputs from safety sensors and the recirculation subsystems to exert overarching control over the recirculation process to maintain the safety of the operators and the facility in case of any hazardous deviations from normal operating conditions. Fig.~\ref{fig:slow_control} shows the concept of the 30-ton slow control system.

The central process control is primarily required to manage the flow of liquid into and out of the main detector tank by controlling the operational state of the two recirculation centrifugal pumps. Due to variation in the flow resistance through filter housings as well as the operation of subsystems, there will always be an imbalance between the flow into and out of secondary process tanks. Over time, the liquid level in the secondary tanks changes and therefore, the PLC monitors the tank liquid levels and signals the upstream or downstream equipment to start or stop pumps as needed to balance the tank liquid levels. 
The central control system has also been designed with the intention of providing a human-machine interface (HMI) for the operators to interact with the detector recirculation system. The control panel provides a tactile means for a human operator to assume manual control under the supervision of the global safety control system and, under certain circumstances, override certain safety interlocks. The HMI also includes features for visualizing the real-time process parameter data measured by the sensors and logging the data for historical data visualization and offline data analysis.

\subsection{Control System Hardware}
As with most process control systems, the hardware of the central control system and the global safety system are structured in layers. The field device layer consists of input and output devices such as sensors, switches, and relays. The controller layer consists of a PLC for process control and a safety relay for safety oversight. The supervisory monitoring layer consists of instrumentation and hardware for the user interface and data logging. The field devices are located in specific equipment and parts of the recirculation system. The controller and HMI are in a control panel mounted at a central location in the facility near the detector recirculation system. The need to interact with the system is only required in an emergency, or when the operation mode is to be changed, or in case of maintenance intervention for the recirculation system components. Run conditions are quite stable, requiring only minimal intervention, and remote monitoring over the BNL intranet through the HMI web server is usually sufficient to check the operating status of the system.

The core of the control system is the PLC, an Allen Bradley CompactLogix 5069 – L320ER, which is a ruggedized industrial computer capable of withstanding harsh environments (temperatures, humidity, dust, electrical noise, etc.) and carry out control functions in real time with precision and reliability. The PLC collects real-time data from various sources, such as the process sensors and feedback signals from the equipment via input and output (I/O) modules that communicate with the PLC using the backplane or using the IO-Link protocol. Process sensors transmit measured values as analog 4 – 20 mA current signals, while feedback inputs from equipment and certain safety sensors transmit digital (ON or OFF) signals. The Allen-Bradley Studio 5000 Logix Designer software is used to program the control logic and configuration that the PLC uses to read the input signals and execute the appropriate output commands. 

The safe operation of the recirculation system is managed by a safety relay, which monitors key safety-related parameters and initiates a safe shutdown of the WbLS recirculation system in case of a fault or hazardous situation. The Phoenix Contact safety relay (PSR-SCP- 24UC/ESAM4/8X1/1X2) has a release time of <20 ms according to the technical sheet provided by the manufacturer. The safety relay has two independent and redundant channels that monitor input signals from safety-specific sensors such as emergency stop buttons, leak detectors, tank overfill float switches, pump motor overload power monitors, subsystem fault indicators, etc. The safety relay continuously monitors these channels along with its own internal circuitry to make sure that everything is operating correctly, and if any of these checks fail, then a safe shutdown is triggered. The safety relay has force-guided contacts that function as the enabling channels for control signals from the PLC to the various controlled outputs, and therefore, in the event of an emergency condition, the enabling channels are interrupted, and all controlled outputs are stopped safely.
The field devices integrated with the control system form the various inputs and outputs that provide the real-time operational status of the 30-ton detector recirculation system. Communication between the field devices and the PLC is via analog (4 – 20 mA) current loops, discrete, i.e., ON/OFF voltage signals, as well as EtherNet/IP data networks. Some analog devices are wired directly to the PLC input cards, which read and process their 4 – 20 mA current signals. These so-called current loop sensors receive their operating power via the PLC or directly from an independent power supply. Some sensors are connected to IO-Link modules (Keyence NQ-MP8L), which provide the operating power to the sensors, read and process the signals, and communicate the processed sensor data to the PLC. Discrete sensors and controlled outputs have been wired to communicate with the PLC indirectly via control relays. The low-powered PLC discrete input/output cards are therefore isolated from the high-powered field devices, which provides circuit protection to the PLC from overloads, short circuits, and faults in the field devices. 

Ultrasonic level sensors are used to monitor the liquid level in the detector tank and the secondary process tanks. The sensors are mounted on top of the tanks in threaded ports. A non-contact liquid level sensor (FlowLine EchoPod UG06) is used to measure the liquid level in the main tank, due to its inherent condensation resistance and to avoid any sensor probes being submerged into the WbLS. Ultrasonic probe level sensors (Keyence FL-001) are used to monitor the level in the secondary process tanks, which are used in the control logic by the PLC to determine pump operation states. Clamp-on ultrasonic flow meters (Keyence FD-R series) are used to monitor the flow rate of liquid at the inlet and outlet of the main detector tank. Temperature probes (Keyence FI-T50) are used to monitor the temperature of the liquid flowing into and out of the heat exchanger. This serves to track the temperature of the process liquid as well as the proper operation of the heat exchanger. A pressure transmitter (Omega PX series) serves to monitor the pressure at the outlet of the main circulation pump. The power required by the pumps or pump load is monitored by ProcessDefender load monitors, which are programmable to set off alarms and control relay circuits if the pump power load reaches certain thresholds for safety operations. The pressure in the main tank is maintained within the tank's rated range of $-1$ to $+1$~psi.

Vertical mount float switches attached at the top of each tank serve as discrete sensors for the high-high liquid level and are wired into the safety circuit and to the PLC in parallel to provide a signal for safety shutdown in case of a tank overfill condition. A BAPI leak detector with the rope sensor affixed along the perimeter of the secondary containment provides another safety-related monitoring sensor, which is used to trigger a safety shutdown and alarm condition.

Controlled outputs include a pump motor control circuit (MCC) consisting of soft starters and contractor relays. The MCC can be controlled automatically by the PLC to operate the pumps depending on the tank level readings as discussed above, but it also has a manual override functionality to operate the pumps independently of the PLC inputs if the need arises. Electrically operated control valves are also operated by the PLC in sequence with the pump operation to provide additional means of control of the liquid flowing into and out of the main detector tank. The control valves have a battery backup safety system designed to fail the valves in the closed position in case of a shutdown due to an emergency or loss of power.

\subsection{Control and Monitoring Interface}

\begin{figure}
    \centering
    \includegraphics[width=0.9\textwidth]
 {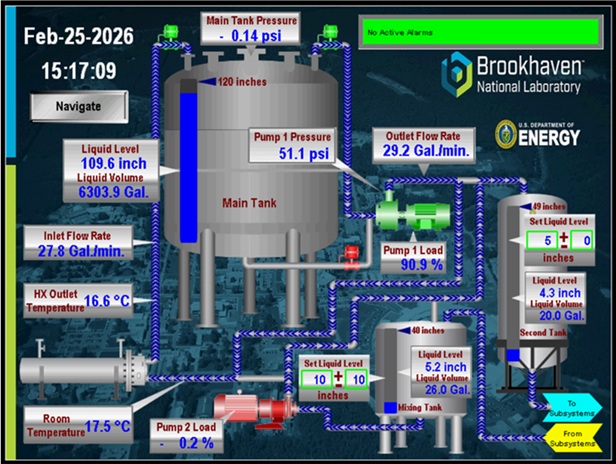} 
    \caption{Process data monitoring interface of the WbLS 30 – ton detector central control system.}
    \label{fig:sc_interface}
\end{figure}

A user interface provides the ability for a user/operator to interact with the control system locally via a control panel that has a touch screen display, as well as remotely over the laboratory intranet. The figure below shows some of the windows and features available for system monitoring and control via the interface. 
The HMI was created using the Crimson 3.2 software which includes a powerful set of built-in graphic tools for creating rich, functional user interfaces.
Fig.~\ref{fig:sc_interface} shows the interface window on the monitor.
The main process monitoring page of the interface provides a visualization of all the main sensor readings as well as the operational state of the controlled outputs. The page has graphics to indicate the liquid circulation path through the system, displays for the sensor readings, indicators that change state (OFF, visible, or flashing) or color (red, yellow, or green, etc.) for displaying different operational conditions or to draw attention to any alarm conditions. The user interface system also has a built-in notification system, which sends an alert via email to operators when certain process conditions are met or an alarm is triggered.

\begin{figure}
    \centering
    \includegraphics[width=0.9\textwidth]
 {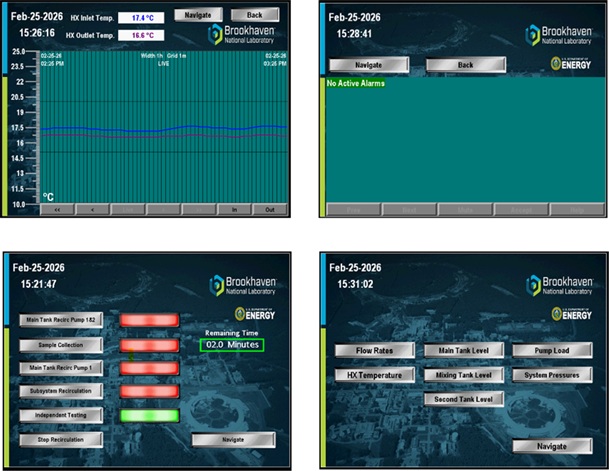} 
    \caption{WbLS 30-ton detector central control system interface. The top left panel shows a window with a plot of temperature sensor readings, the top right panel shows the system alarm viewer, the bottom left panel shows the menu to select between different liquid recirculation paths, and the bottom right panel shows a menu to access a data trend plot for various other process sensors. }
    \label{fig:sc_interface2}
\end{figure}

The user can navigate the interface system to reach menus and pages for controlling and monitoring different parameters of the system. The interface system locally stores sensor data over the previous seven days, and the data trend screens can visualize the data plots over this time period. The sensor data logs stored by the system older than seven days are automatically downloaded to a PC for storage, offline data processing and visualization. 
Fig.~\ref{fig:sc_interface2} shows several user interface indicators in the monitor.

\section{Commissioning}  
\label{sec:commisioning}

The commissioning of the 30-ton detector was an intensive, multi-month campaign to integrate and stabilize all hardware and software components before the introduction of the WbLS. This entire period, from the initial power-on to the start of stable operations, was conducted with the tank filled with purified water, providing a crucial baseline for detector performance. The goal is to establish a trigger scheme capable of tagging muons that traverse the tank. These tagged muon events will serve as a proxy for monitoring the detector stability over time.

\begin{figure}[htbp]
\begin{center}
\includegraphics[width=1\textwidth,trim=0 0 0 0,clip]{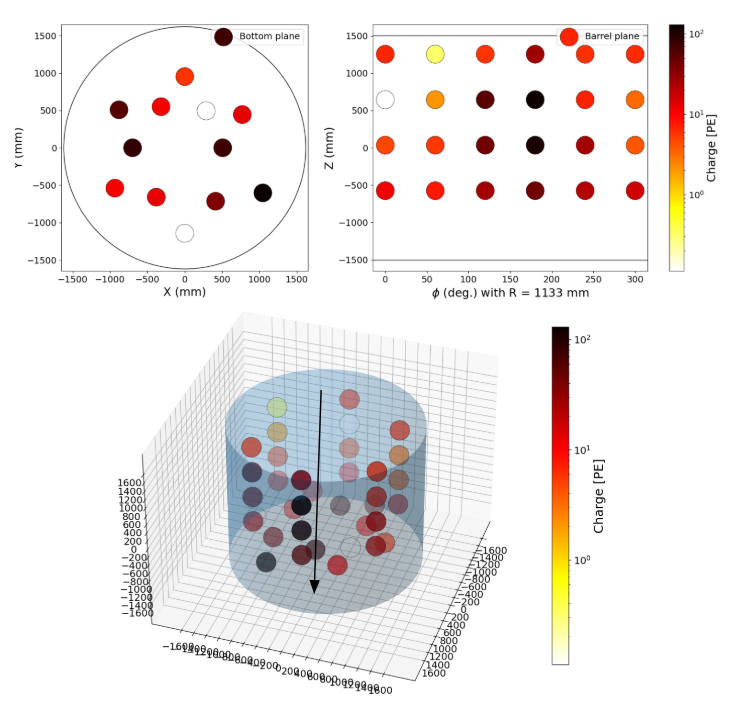}
\caption{\label{fig:event_display} Event display for a cosmic muon candidate. The top two panels show the bottom and wall views of the PMT hits. The bottom panel shows the 3D view. Each circle corresponds to one PMT. The color scale of the circle represents the charge in the PMT. Three inactive PMTs have been removed from the event display. The purpose is to show the expected ring shape pattern in the 30-ton detector with full PMT charge responses.}
\end{center}
\end{figure}

The process officially began on July 23, 2024, with the first activation of the PMTs. 
The trigger is formed by multi-layer scintillator paddles, with two layers of paddles installed above the tank and two layers below. Each layer is composed of four individual 20 cm x 90 cm plastic scintillator bars, arranged to cover the central region of the detector. To streamline the system, the four paddles within each layer share a common high-voltage supply. Their signals are processed through discriminators to create uniform logic pulses, which are then timed and recorded by a CAEN V1740 digitizer. A significant effort was made to tune each individual paddle to a steady rate of approximately 20 Hz, ensuring high and consistent trigger efficiency.
In order to tag the crossing muons, we require a time coincidence between signals in the top and bottom paddle layers as well as a majority signal from at least five of the bottom PMTs.
A second, independent trigger path was established for the alpha source calibration system, allowing for dedicated low-light calibration runs without interference from the cosmic ray flux.
Fig.~\ref{fig:event_display} shows a cosmic muon candidate display in an early run demonstrating that all the PMTs were functioning as expected during the commissioning.

In the final phase of commissioning, the detector's vital support systems were validated, including comprehensive tests of the WbLS pumping and circulation system to ensure its readiness for long-term operation.
By January 2025, all detector subsystems—the trigger, high voltage, DAQ, calibration, and preliminary circulation systems—were fully integrated, tested, and stabilized. The commissioning phase was declared complete, and the detector began a prolonged period of stable data-taking with water. This steady-state operation provided a high-quality baseline dataset, essential for understanding the detector's response before the first injection of WbLS concentrate on April 1, 2025.

\section{WbLS Production and Injection} 
\label{sec:injection}
Following stable operation during the pure-water phase, WbLS injection was initiated on April 1, 2025. The WbLS mixture was prepared at the Brookhaven National Laboratory’s Liquid Scintillator Production Facility (LSPF), which has been addressing the consistency challenges commonly faced by chemical manufacturers in maintaining product purity across production batches. Notably, it is the only ton-scale liquid scintillator production facility at an academic institution in the U.S.
 

\subsection{Purification and Sequential-Mixing Workflow}

To produce high-purity WbLS, raw materials undergo multiple purification steps, including filtration, vacuum distillation, ion-exchange chromatography, pH adjustment, and recrystallization, targeting the removal of free ions, color-quenching impurities, and volatile components. Notably, the LSPF uses a pilot-scale short-path thin-film vacuum distillation system\footnote{For more information, please see:~\href{LSPF}{https://www.bnl.gov/chemistry/nnc/ls-production-facility.php}}, capable of processing up to 40 kg per hour under $\approx 10^{-4}$ mbar vacuum, with staged low-temperature condensation to collect the volatile byproducts and improve the optical clarity.

The purified materials are then transferred into a 1,000-liter sequential mixing reactor, where synthesis proceeds under an inert atmosphere. Flow and reaction control are tightly managed to ensure high-quality batch output. Post-synthesis, the product undergoes rigorous QA/QC testing, including—but not limited to—measurements via Ultraviolet (UV)/Visible (Vis) spectrophotometry, Liquid Chromatography--Mass Spectrometry (LC-MS), Gas Chromatography--Mass Spectrometry (GC-MS), X-ray Fluorescence (XRF), and attenuation tests.

\subsection{Transportation and Injection Protocol}

The synthesized WbLS organic component is stored and transported in custom-lined 55-gallon steel drums, which reduce oxygen and impurity ingress during transit. At the detector site, the system’s two-pump circulation loop is activated 24 hours prior to injection. Parameters such as high voltage, DAQ, temperature, pressure, and flow rate are carefully monitored to ensure detector stability.

\begin{figure}[htbp]
\begin{center}
\includegraphics[width=.49\textwidth,trim=0 0 0 0,clip]{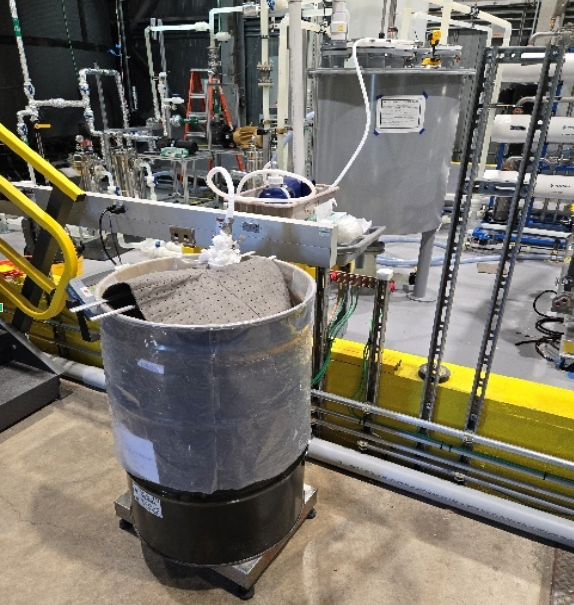}
\centering 
\includegraphics[width=.49\textwidth,trim=0 0 0 0,clip]{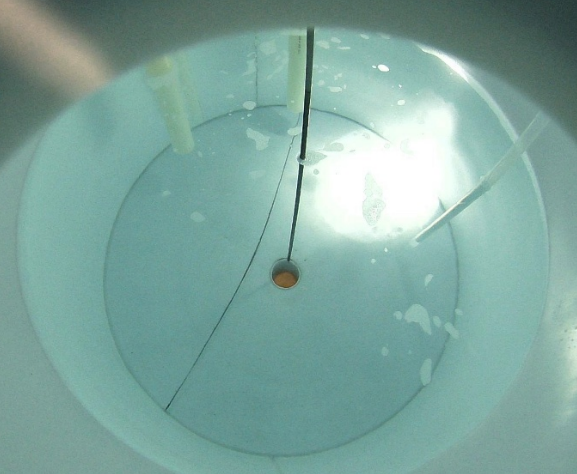}
\caption{\label{fig:injection_rr} Left: injection assembly. Right: mixing tank during the injection. }
\end{center}
\end{figure}

The WbLS is then metered from the drum into the mixing tank using a precision metering pump. The target concentration in the 30-ton detector tank is 1\% WbLS by mass, achieved through three staged injections at 0.35\%, 0.75\%, and 1\% by mass, each administered over approximately seven hours.
In Fig.~\ref{fig:injection_rr}, the left panel shows the chemical assembly for the injection and the right panel shows the mixing tank liquid during the actual injection. As mentioned at the beginning of this section, the first injection, bringing the WbLS concentration to 0.35\%, was performed on April 1, 2025. This was followed by a second injection, increasing the concentration to 0.75\%, on April 23, 2025, and a final injection, reaching 1\%, on May 7, 2025.

\section{Detector Calibration} 
\label{sec:calibration}

Following the procedure established for the 1‑ton prototype \cite{1ton-paper}, the PMT gain ($G$) of the 30-ton detector is determined from the SPE peak, which appears when the average number of photoelectrons (PE) produced at the photocathode is less than one. This SPE-based calibration provides a direct model‑independent conversion from the integrated anode charge to photoelectrons.

We performed PMT gain calibration with a low‑intensity $^{210}$Pb radioactive source positioned at the center of the detector. The source assembly, identical to that used for the 1‑ton prototype, consists of a sealed needle embedded in an EJ‑228 plastic scintillator. The plastic scintillator is a cylindrical rod, 5~mm in diameter and 25~mm in length, viewed by a 10~mm-diameter PMT (Hamamatsu R1635) housed inside an acrylic tube.
The emissions that are most relevant for our calibration from the $^{210}$Pb decay scheme \cite{1ton-paper} are the 1161 keV $\beta$-particle (CSDA range $\approx$ 5 mm) and the 5305 keV $\alpha$-particle (CSDA range $\approx$ 40 $\mu$m). Both particles deposit their energies within the EJ-228 plastic, producing a light spectrum with a yield of roughly 10,000 photons/MeV for the minimum-ionizing $\beta$ particles and about 1000 photons/MeV for $\alpha$ particles, in which the latter is reduced by the quenching effect. The remaining $\beta$ and $\gamma$ emissions contribute negligibly to both the trigger rate and the total light yield. A fraction of this light reaches the small R1635 PMT, generating the start signal for data acquisition. We set the discriminator to achieve a 20~Hz raw trigger rate, then prescaled to $\sim$1~Hz to maintain an acceptable data throughput. Most scintillation photons exit the plastic and acrylic housing, illuminating the 30-ton detector nearly isotropically.

\begin{figure}[htbp]
\begin{center}
\includegraphics[width=.49\textwidth,trim=0 0 0 0,clip]{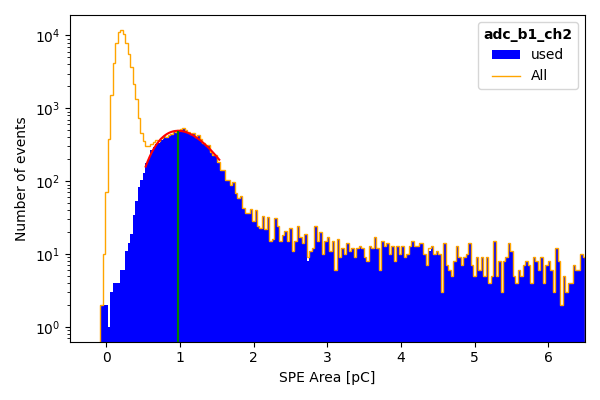}
\centering 
\includegraphics[width=.49\textwidth,trim=0 0 0 0,clip]{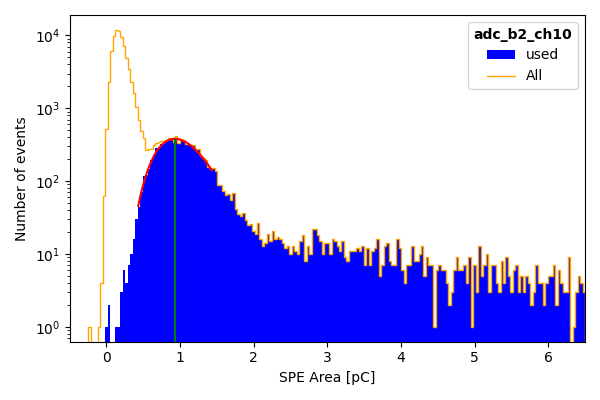}
\caption{\label{fig:spe_peak_before_alpha} Charge distribution of SPE events fitted with a Polya function from data collected on 15 April 2025. Left: bottom PMT (channel adc\_b1\_ch2); right: side PMT (channel adc\_b2\_ch10). The orange histogram includes all waveforms collected during the calibration runs, while the blue histogram represents only the waveforms selected for calibration. The fitted peak, centered around 1~pC, corresponds to the SPE peak.}
\end{center}
\end{figure}

On April 24 2025, water leaked into the acrylic calibration tube, causing the $\alpha$-source tag PMT to fail. After this event, we adopted an unbiased random trigger method for calibration. A NIM pulser delivered triggers at 300~Hz, and for each trigger, we recorded a 4~$\mu$s time window from every channel. A typical 10-minute run accumulated sufficient single-PE waveforms for each PMT. After the baseline subtraction, we selected the first pulse in each waveform and integrated its charge. For the selection, we removed the pedestal by using threshold based selection; selecting only pulses which has threshold greater than 2.5 mV. The charge around zero pC is reflecting very low height pulses after selection.  The resulting charge distribution was fitted with a Polya function~\cite{polya-method}. The SPE mean obtained from the fit agreed with previous results from the $\alpha$-source-tagged method, demonstrating continuity between calibration approaches.
Figures~\ref{fig:spe_peak_before_alpha} 
show the SPE charge distribution for a side and a bottom PMT.

After the source PMT failure, two photomultiplier channels went offline: a bottom PMT lost high-voltage stability on April 28 2025, and a side PMT failed in the same way on June 03 2025. Another bottom PMT channel had failed previously. All three channels were excluded from subsequent gain calibrations and light-yield analyses.

\begin{figure}[htbp]
\begin{center}
\includegraphics[width=.49\textwidth,trim=0 0 0 0,clip]{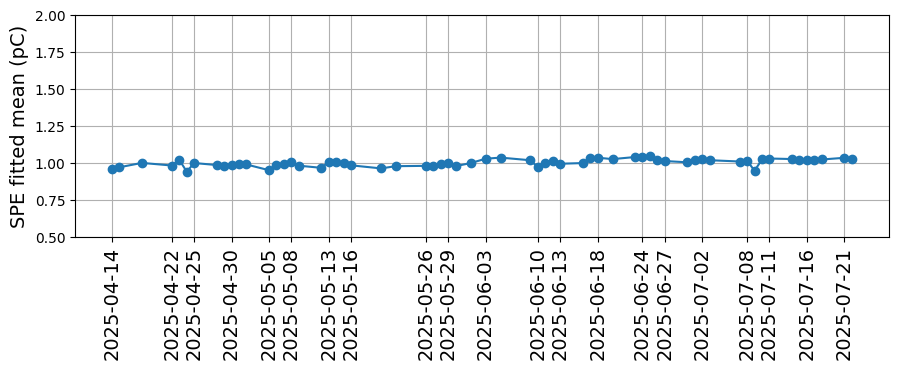}
\centering 
\includegraphics[width=.49\textwidth,trim=0 0 0 0,clip]{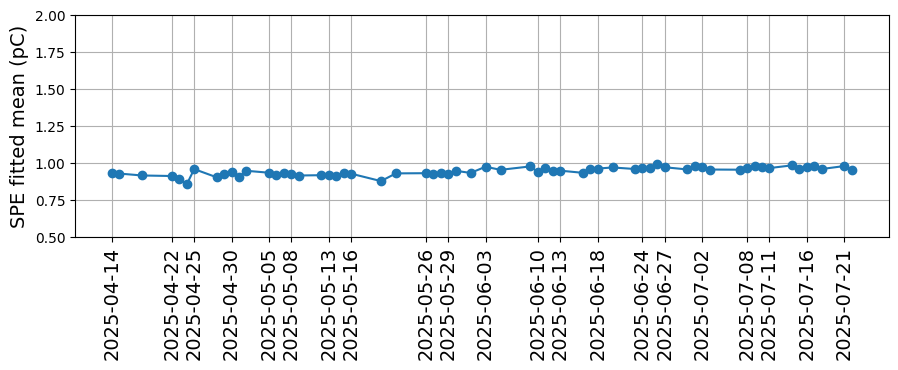}
\caption{\label{fig:30t_spe_stability} Daily fitted SPE mean charge before and after the $\alpha$ source PMT failure. Left: bottom PMT (channel b1\_ch2); right: side PMT (channel b2\_ch10). The SPE mean value ($\sim$1~pC) indicates stable operation during data-taking.}
\end{center}
\end{figure}

\begin{figure}[htbp]
\begin{center}
\includegraphics[width=.49\textwidth,trim=0 0 0 0,clip]{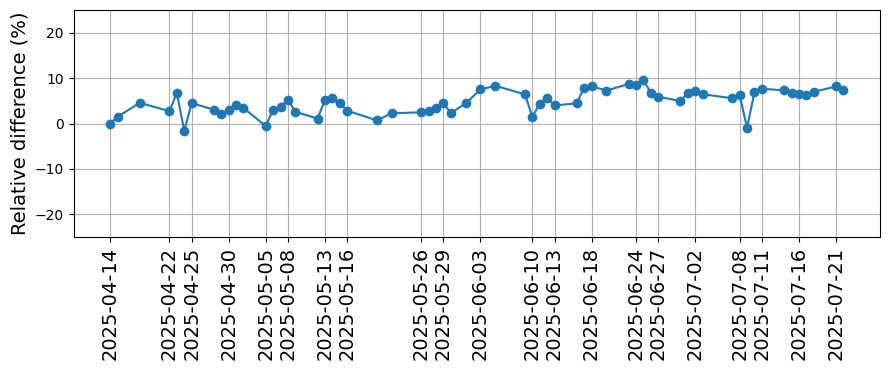}
\centering 
\includegraphics[width=.49\textwidth,trim=0 0 0 0,clip]{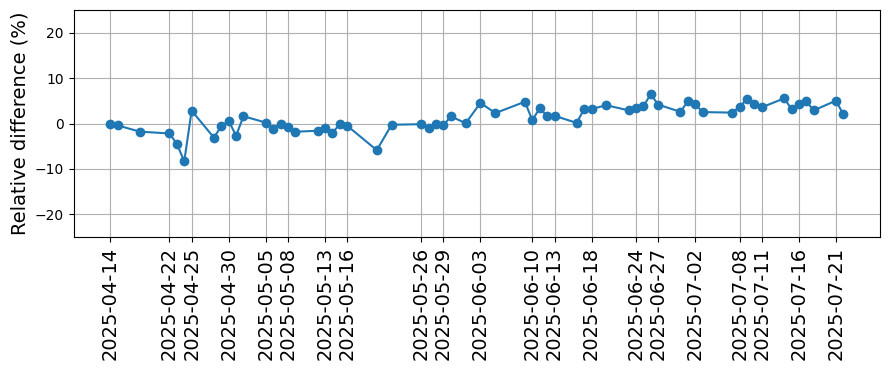}
\caption{\label{fig:relative_difference} Daily relative percentage difference of the fitted SPE mean charge before and after the $\alpha$ source outage. Left: bottom PMT, channel b1\_ch2; right: side PMT, channel b2\_ch10. Relative differences remain under 10\%, indicating stable operation during data-taking.}
\end{center}
\end{figure}

A key goal was to quantify detector stability during the transition from pure water to 1\%~(by mass) WbLS, injected on May 7, 2025, and throughout an extended operation period. We computed the SPE mean charge for each active channel over the interval from April 14, 2025, to July 21, 2025, including both the WbLS injection and the $\alpha$-source PMT failure. The detector response increased rapidly during mixing and stabilized within a few hours; thereafter, daily variations were limited to a few percent. The daily SPE mean charge for representative side and bottom PMTs before and after the $\alpha$-source PMT failure are shown in Figure~\ref{fig:30t_spe_stability}. A nearly constant  SPE mean charge ($\sim$ 1~pC) was maintained throughout the operation, demonstrating stable PMT performance. Figure~\ref{fig:relative_difference} presents the relative percentage difference of SPE mean charges measured before and after the $\alpha$ source outage. The variation remained under 10\%, further demonstrating the stability of PMTs throughout the operation. Gain monitoring was continued daily to ensure stable operation of both the PMTs and their high-voltage supplies.

\section{Preliminary Performance} 
\label{sec:performance}
\begin{figure}[htbp]
\begin{center}
\includegraphics[width=0.8\textwidth,trim=0 0 0 0,clip]{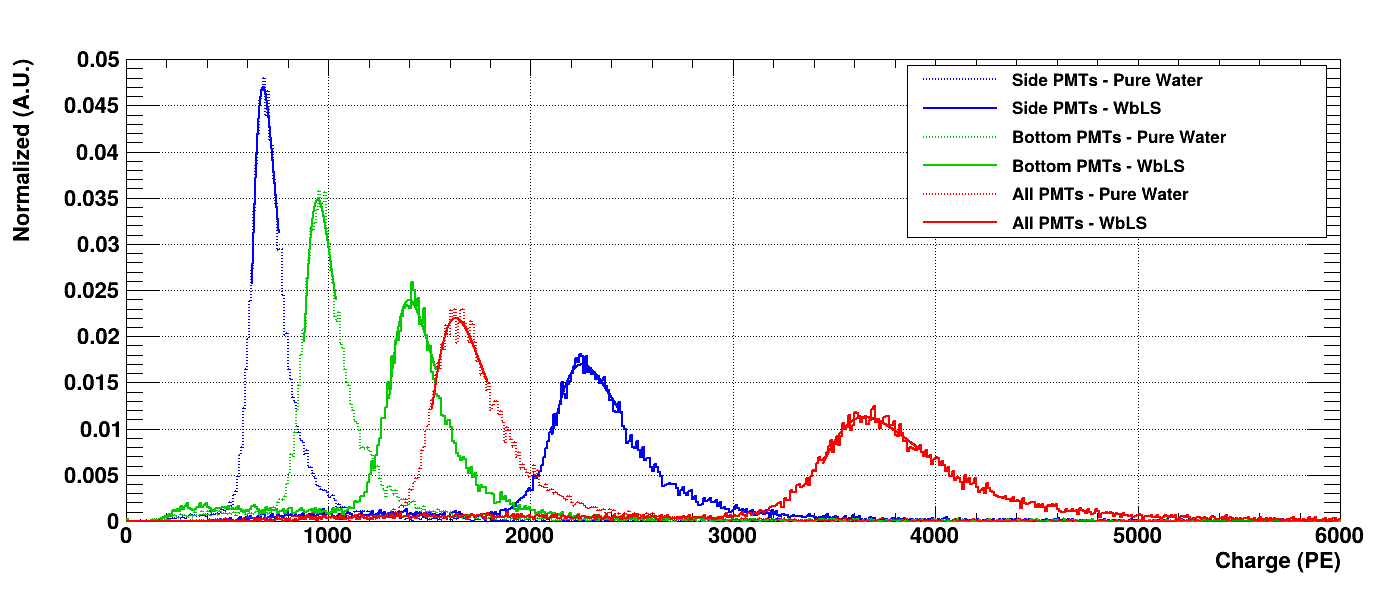}
\caption{\label{fig:crossing_muon_pmts} Charge distributions for the side, bottom, and all PMTs. The dashed histograms show the pure-water data, while the solid histograms show the 1\% WbLS data. Each histogram is normalized to unit area.
.}.
\end{center}
\end{figure}

Crossing muons were used to monitor the detector performance during the pure-water run and throughout the injection period, and are also used to track the stability of the WbLS after injection. This method demonstrates the stability of the detector performance during operation with pure water and provides a framework for monitoring the WbLS stability after injection. To select these events, the data analysis required that the bottom scintillator paddles be triggered in coincidence with the top paddle trigger. Fig.~\ref{fig:crossing_muon_pmts} shows an example of the crossing muon peaks observed in all, side and bottom PMTs. Two bottom PMTs (B1ch1 and B1ch9) and one side PMT (B2ch8) were inactive during this period, and these channels were excluded from the subsequent analysis. Each peak was fitted with a Landau function to extract the peak light yield, which was then used to monitor the stability of the 30T detector. 

Table~\ref{tab:mpv_ratio} summarizes the most probable value (MPV) obtained after landau fit for side, bottom and all PMT histograms. The larger enhancement observed in the side region relative to the bottom region indicates a spatial dependence of the WbLS light-yield, reflecting differences in photocathode coverage and light-collection geometry between the two regions.

\begin{table}[htbp]
\centering
\caption{Landau fit results for the most probable value (MPV) of the
photoelectron (PE) distributions, with the light-yield gain given as the
WbLS (1\%)-to-pure-water MPV ratio for each group.}
\label{tab:mpv_ratio}
\begin{tabular}{llcc}
\hline
\hline
Group  & Run      & MPV (PE)          & MPV ratio (WbLS/Water) \\
\hline

\multirow{2}{*}{Side}   & Pure Water & $688.15 \pm 1.39$    & \multirow{2}{*}{$3.31 \pm 0.01$} \\
                        & WbLS       & $2278.52 \pm 3.98$   & \\
\hline
\multirow{2}{*}{Bottom} & Pure Water & $960.08 \pm 1.92$    & \multirow{2}{*}{$1.48 \pm 0.00$} \\
                        & WbLS       & $1416.65 \pm 2.70$   & \\
\hline
\multirow{2}{*}{All}    & Pure Water & $1650.20 \pm 3.21$  & \multirow{2}{*}{$2.23 \pm 0.01$} \\
                        & WbLS       & $3685.40 \pm 5.72$  & \\

\hline
\hline
\end{tabular}
\end{table}

Regular data taking continued throughout the injection, allowing the detector’s transition behavior to be observed in real time. The injection lasted approximately two hours and was deliberately performed slowly to ensure sufficient statistics for analysis. The PMT array was configured such that, in the pure water phase, the top two rows of side PMTs were positioned outside the Cherenkov cone. Consequently, once LS was introduced into the tank, a substantial increase in light yield was expected on the side PMTs. To study this transition, we selected muon events that passed through the top scintillator paddles and triggered the majority of PMTs, enabling statistically meaningful analysis on a minute-by-minute timescale. After the transition period, the analysis shifted to using crossing muons to monitor the detector’s long-term stability.

\begin{figure}[htbp]
\begin{center}
\includegraphics[width=0.8\textwidth,trim=0 0 0 0,clip]{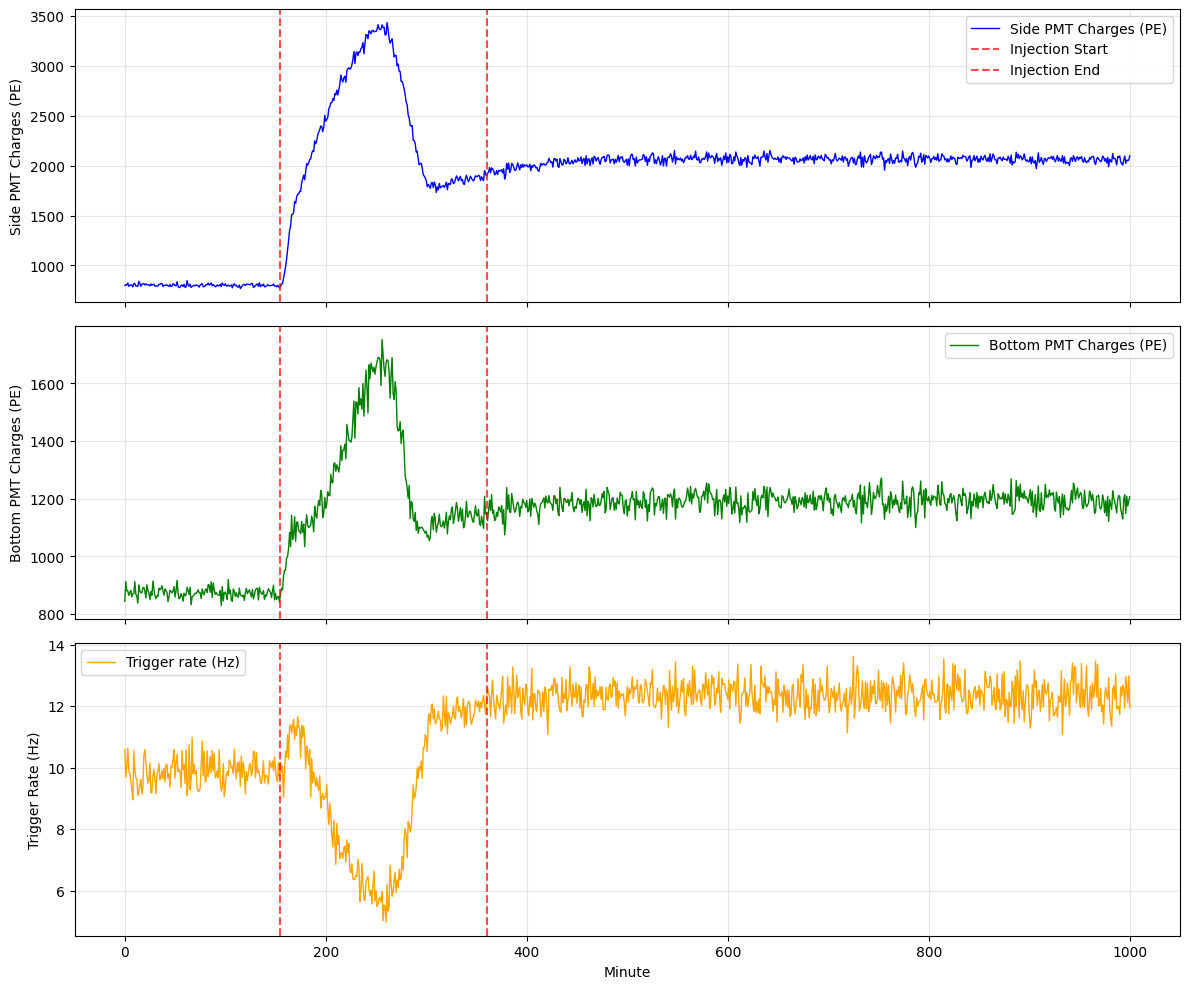}
\caption{\label{fig:injection_transition}Truncated mean charge of the side PMTs (top panel) and bottom PMTs (middle panel) of the 30T demonstrator during the liquid scintillator (LS) injection. The trigger rate is shown in the bottom panel. The two dashed red lines indicate the start and end times of the injection. See the text for further explanation.}
\end{center}
\end{figure}

Fig.~\ref{fig:injection_transition} shows the light yield measurements during the injection process, separated for the side and bottom PMTs. For each group, the light yield was calculated using the truncated mean to suppress the influence of occasional large or small events that could bias the minute-by-minute averages. The injection from pure water to 0.35\% LS exhibits an intriguing pattern: the light yield rises sharply at first, reaches a peak, then drops, and finally increases again gradually to its final level. 

This behavior was further investigated using Monte Carlo simulations, in which a highly scattering region was introduced to represent the dense LS region. The observed behavior was found to be correlated with both the trigger logic and the mixing state of the LS within the tank. When LS micelles were initially injected, they did not disperse uniformly. Instead, a "micelle-dense" region formed near the injection point, a highly scattering zone, which reflected Cherenkov light away from the bottom PMTs.
As a result, fewer bottom PMTs were hit than usual, often falling short of the trigger requirement. This effect is clearly seen in the drop in trigger rate per minute shown in Fig.~\ref{fig:injection_transition}. Initially, as micelles began entering the tank, the added scintillation light boosted both the trigger rate and the observed light yield. However, once the micelle-dense region became more concentrated, the trigger rate dropped. During this phase, only muons passing directly through the dense region could produce enough light to satisfy the trigger, resulting in fewer, but brighter events. As more micelles were injected and the tank’s circulation system began to break up the concentrated region, the light yield peaked and then started to decline. At the same time, the trigger rate gradually recovered, indicating more uniform coverage. Eventually, once the LS micelles became evenly mixed throughout the detector volume, both the light yield and trigger rate stabilized at their expected higher levels. 

\begin{figure}[ht]
    \centering
    
    \begin{subfigure}[b]{0.9\textwidth}
        \centering
        \includegraphics[width=\textwidth]{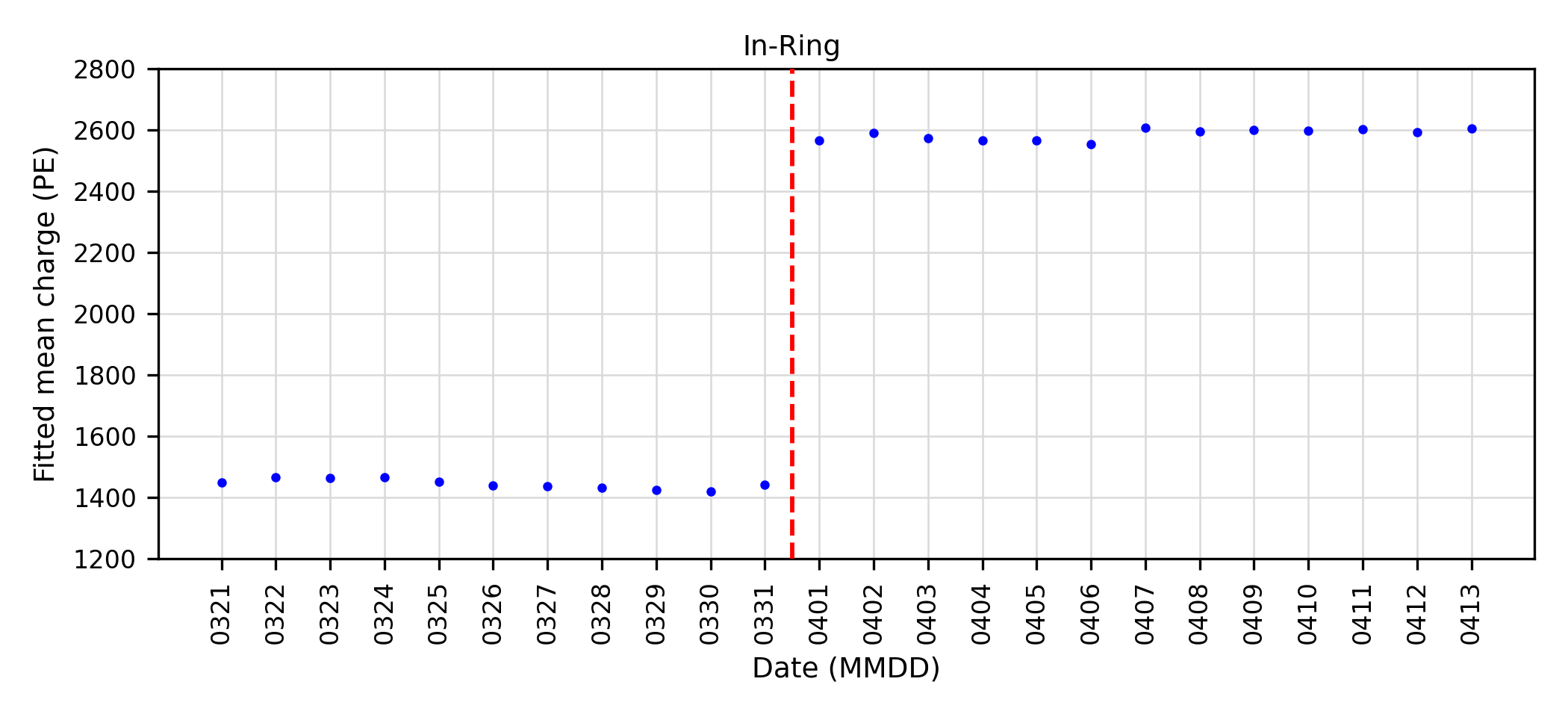}
        \caption{In-ring PMTs}
        \label{fig:1a}
    \end{subfigure}
    \hfill
    \begin{subfigure}[b]{0.9\textwidth}
        \centering
        \includegraphics[width=\textwidth]{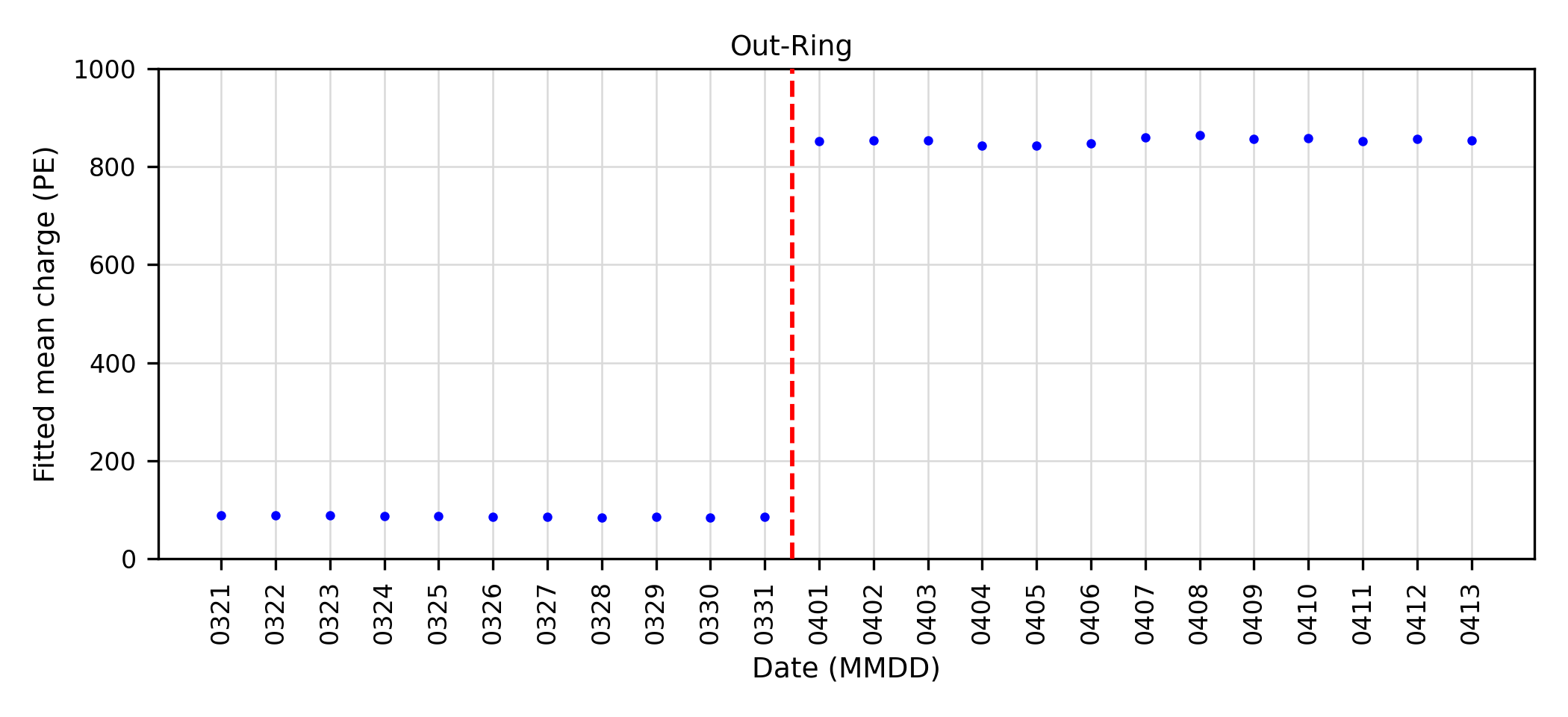}
        \caption{Out-ring PMTs}
        \label{fig:1b}
    \end{subfigure}

    \caption{Preliminary mean photoelectron yield of crossing muons as a function of date for (a) in-ring PMTs and (b) out-ring PMTs, shown before the injection (pure water) and after the WbLS injection (at 0.35\% by mass). The red dashed vertical line marks the injection day.}
    \label{fig:inandout}
\end{figure}

These phenomena, supported by both data analysis and Monte Carlo simulation, highlight the capability of the 30T demonstrator in terms of both detector design and analysis framework. 
A quantitative data--simulation comparison of the injection transient will be included in the forthcoming simulation publication.
We are actively monitoring the detector response's long-term stability through continuous calibration and source-based monitoring routines.
The preliminary WbLS performance, in particular the light yield, is traced over time with the crossing muon samples as shown in Fig.~\ref{fig:inandout}. The system has operated stably during the pure water phase for several months and for several days immediately following the injection. The stability has been continuously monitored, and the long-term stability of the WbLS runs will be presented in a future publication.
In addition, environmental data such as temperature, humidity, and circulation flow are being recorded along with the detected light in the detector. Correlation between the environment and the detected light in the detector can be measured in order to assess their influence on the data stability. For example, higher temperatures generally result in lower light yield, and vice versa~\cite{XIE2021165459}. Since the temperature of the system has remained  stable, the impact of temperature fluctuations on the measured light yield is expected to be minimal.
The environmental effects can be combined quantitatively, allowing us to predict and compensate for future fluctuations in detector response. 
A comprehensive Monte Carlo simulation framework employing RATPAC-TWO~\cite{ratpac_two} for the 30T demonstrator has been developed and benchmarked against experimental data. A detailed analysis of the results will be presented in a forthcoming publication. The next milestone in this program involves gadolinium doping in the WbLS, coupled with the deployment of a novel AmBe calibration system integrated with an LYSO tagging module. This configuration will enable precise neutron–gamma coincidence measurements, providing a powerful tool to study neutrino-like interactions and verify the uniformity of the gadolinium-doped WbLS mixture.  

\section{Conclusion} 
\label{sec:conclusion}

This paper has detailed the successful design, construction, commissioning, and initial operation of a 30-ton Water-based Liquid Scintillator demonstrator at Brookhaven National Laboratory. This facility represents a crucial, order-of-magnitude scale-up from previous 1-ton demonstrators and serves as an essential intermediate step toward future kiloton-scale experiments. We have demonstrated the stable operation of all key detector subsystems, including the 36-PMT array, the data acquisition and trigger systems, the preliminary liquid circulation and purification plant, and the comprehensive slow control system. The integration of the nanofiltration and Gd-loading systems, together with an evaluation of the resulting detector performance, will be presented in a future publication.

The commissioning phase, conducted with Omega-18 water, established a stable operational baseline and validated the performance of the cosmic muon tagging system. The subsequent PMT calibration campaign confirmed the long-term gain stability of the photosensors, which was successfully maintained using both a source-tagged method and a random trigger method after the source tagger failed. The successful production of high-purity WbLS concentrate and its injection into the detector marked a major milestone. Initial performance results using cosmic muon data clearly show the expected increase in light yield and provide unique insights into the liquid's mixing dynamics during the transition from water to WbLS. The detector is now operating stably, providing valuable data on the long-term behavior of WbLS at an unprecedented scale. We plan to first complete initial long-term light-yield monitoring using cosmic muons, followed by commissioning of the liquid circulation and purification system, including nanofiltration and the Gd-water system, and finally full system-integration testing.

The successful operation of the 30-ton prototype is a significant achievement for the WbLS R\&D program. The experience gained and the data collected are invaluable for de-risking the designs of future large-scale detectors and demonstrating the viability and scalability of WbLS technology for a broad range of physics applications. Detailed analyses of the detector performance and comparisons with our comprehensive Monte Carlo simulations are ongoing and will be the subject of future publications.

\acknowledgments


The work conducted at Brookhaven National Laboratory was supported by the U.S. Department of Energy under contract DE-AC02-98CH10886. 
Work conducted at Lawrence Berkeley National Laboratory was performed under the auspices of the U.S. Department of Energy under Contract DE-AC02-05CH11231.  The project was funded by the U.S. Department of Energy, National Nuclear Security Administration, Office of Defense Nuclear Non-proliferation Research and Development (DNN R\&D).  
This material is based upon work supported by the U.S. Department of Energy, Office of Science, Office of High Energy Physics, under Award Numbers DE-SC0018974, DE-SC0012704 and DE-SC0012447. 
Work conducted at Chung-Ang University was supported by the National Research Foundation of Korea through grant NRF-2022R1A2C1009686.





\bibliographystyle{JHEP}
\bibliography{biblio_fixed}

@article{1ton-paper,
    author  = "Xiang, X. and others",
    title   = "{Design, construction, and operation of a 1-ton Water-based Liquid scintillator detector at Brookhaven National Laboratory}",
    journal = "Journal of Instrumentation",
    volume  = "19",
    number  = "06",
    pages   = "P06033",
    year    = "2024",
    doi     = "10.1088/1748-0221/19/06/P06033"
}

@article{polya-method,
    author  = "Prescott, J. R.",
    title   = "{A statistical model for photomultiplier single-electron statistics}",
    journal = "Nucl. Instrum. Meth. A",
    volume  = "39",
    pages   = "173-179",
    year    = "1966",
    doi     = "10.1016/0029-554X(66)90026-9"
}

@article{Richards:2019lkb,
    author = "Richards, Benjamin",
    editor = "Forti, A. and Betev, L. and Litmaath, M. and Smirnova, O. and Hristov, P.",
    title = "{The ToolDAQ DAQ Software Framework {\&} Its Use In The Hyper-K {\&} ANNIE Detectors}",
    doi = "10.1051/epjconf/201921401022",
    journal = "EPJ Web Conf.",
    volume = "214",
    pages = "01022",
    year = "2019"
}

@article{ANNIE,
doi = {10.1088/1748-0221/19/05/P05070},
url = {https://doi.org/10.1088/1748-0221/19/05/P05070},
year = {2024},
month = {may},
publisher = {IOP Publishing},
volume = {19},
number = {05},
pages = {P05070},
author = {Ascencio-Sosa et al., M. and The ANNIE collaboration},
title = {Deployment of Water-based Liquid Scintillator in the Accelerator Neutrino Neutron Interaction Experiment},
journal = {Journal of Instrumentation}
}

@article{EoS,
doi = {10.1088/1748-0221/18/02/P02009},
url = {https://doi.org/10.1088/1748-0221/18/02/P02009},
year = {2023},
month = {feb},
publisher = {IOP Publishing},
volume = {18},
number = {02},
pages = {P02009},
author = {Anderson et al.,T.},
title = {Eos: conceptual design for a demonstrator of hybrid optical detector technology},
journal = {Journal of Instrumentation}
}

@article{Theia,
  author  = {Askins, M. and Bagdasarian, Z. and Barros, N. and others},
  title   = {Theia: an advanced optical neutrino detector},
  journal = {Eur. Phys. J. C},
  year    = {2020},
  volume  = {80},
  pages   = {416},
  doi     = {10.1140/epjc/s10052-020-7977-8},
}

@misc{ratpac_two,
  collaboration = {EoS},
  title         = {Ratpac-two: A Simulation and Analysis Package Based on {Geant4} and {ROOT}},
  howpublished  = {\url{https://github.com/rat-pac/ratpac-two}},
  year          = {2023},
  note          = {Software repository}
}

@article{egads,
    author = "Marti, Ll. and others",
    title = "{Evaluation of gadolinium{\textquoteright}s action on water Cherenkov detector systems with EGADS}",
    eprint = "1908.11532",
    archivePrefix = "arXiv",
    primaryClass = "physics.ins-det",
    doi = "10.1016/j.nima.2020.163549",
    journal = "Nucl. Instrum. Meth. A",
    volume = "959",
    pages = "163549",
    year = "2020"
}

@article{xenonnt,
    author = "Aprile, E. and others",
    collaboration = "XENON",
    title = "{The neutron veto of the XENONnT experiment: results with demineralized water}",
    eprint = "2412.05264",
    archivePrefix = "arXiv",
    primaryClass = "physics.ins-det",
    doi = "10.1140/epjc/s10052-025-14105-0",
    journal = "Eur. Phys. J. C",
    volume = "85",
    number = "6",
    pages = "695",
    year = "2025"
}

@article{coleman,
    title = {Transparency of 0.2\% GdCl3 doped water in a stainless steel test environment},
    journal = {Nucl. Instrum. Meth. A},
    volume = {595},
    number = {2},
    pages = {339-345},
    year = {2008},
    issn = {0168-9002},
    doi = {https://doi.org/10.1016/j.nima.2008.06.049},
    url = {https://www.sciencedirect.com/science/article/pii/S0168900208009108},
    author = {W. Coleman and A. Bernstein and S. Dazeley and R. Svoboda}
}

@article{gwon2025measurementlightyieldresponse,
  title = {Measurement of the light yield response of the Gd-compatible water-based liquid scintillator with the Brookhaven one-ton testbed},
  author = {Gwon, S. and Askins, M. and Asner, D. M. and Baldoni, A. and Cowen, D. F. and Prerez, R. Diaz and Diwan, M. V. and Gokhale, S. and Hans, S. and Kumar, P. and Lawley, G. and Linden, S. and Gann, G. D. Orebi and Park, J. and Reyes, C. and Rosero, R. and Siyeon, K. and Smiley, M. and Wang, J. J. and Wilking, M. and Yang, G. and Yeh, M.},
  journal = {Phys. Rev. D},
  volume = {113},
  issue = {9},
  pages = {092015},
  numpages = {12},
  year = {2026},
  month = {May},
  publisher = {American Physical Society},
  doi = {10.1103/m1qd-yg4r},
  url = {https://link.aps.org/doi/10.1103/m1qd-yg4r}
}

@article{YEH201151,
title = {A new water-based liquid scintillator and potential applications},
journal = {Nucl. Instrum. Meth. A},
volume = {660},
number = {1},
pages = {51-56},
year = {2011},
issn = {0168-9002},
doi = {https://doi.org/10.1016/j.nima.2011.08.040},
url = {https://www.sciencedirect.com/science/article/pii/S0168900211016615},
author = {M. Yeh and S. Hans and W. Beriguete and R. Rosero and L. Hu and R.L. Hahn and M.V. Diwan and D.E. Jaffe and S.H. Kettell and L. Littenberg}
}

@article{Steiger_2024,
doi = {10.1088/1748-0221/19/09/P09008},
url = {https://doi.org/10.1088/1748-0221/19/09/P09008},
year = {2024},
month = {sep},
publisher = {IOP Publishing},
volume = {19},
number = {09},
pages = {P09008},
author = {Steiger, Hans Th.J. and Böhles, Manuel and Stock, Matthias Raphael and Wurm, Michael and Dörflinger, David and Fahrendholz, Ulrike and Mpoukouvalas, Anastasia and Oberauer, Lothar and Steiger, Andreas and Zundel, Dorina},
title = {Development, characterization and production of a novel water-based liquid scintillator based on the Surfactant TRITON™ X-100},
journal = {Journal of Instrumentation}
}

@article{Bhattacharya2026BUTTON30,
  author       = {Bhattacharya, D. S. and Bae, J. and Bergevin, M. and others},
  title        = {Design and development of optical modules for the BUTTON-30 detector},
  journal      = {European Physical Journal Plus},
  volume       = {141},
  pages        = {65},
  year         = {2026},
  doi          = {10.1140/epjp/s13360-025-07266-0},
  url          = {https://doi.org/10.1140/epjp/s13360-025-07266-0}
}

@Article{D0MA00055H,
author ="Onken, Drew R. and Moretti, Federico and Caravaca, Javier and Yeh, Minfang and Orebi Gann, Gabriel D. and Bourret, Edith D.",
title  ="Time response of water-based liquid scintillator from X-ray excitation",
journal  ="Mater. Adv.",
year  ="2020",
volume  ="1",
issue  ="1",
pages  ="71-76",
publisher  ="RSC",
doi  ="10.1039/D0MA00055H",
url  ="http://dx.doi.org/10.1039/D0MA00055H"}

@article{Askins2020THEIA,
  author       = {Askins, M. and Bagdasarian, Z. and Barros, N. and others},
  title        = {THEIA: an advanced optical neutrino detector},
  journal      = {European Physical Journal C},
  volume       = {80},
  pages        = {416},
  year         = {2020},
  doi          = {10.1140/epjc/s10052-020-7977-8},
  url          = {https://doi.org/10.1140/epjc/s10052-020-7977-8}
}

@article{BOREXINO:2023ygs,
    author = "Basilico, D. and others",
    collaboration = "BOREXINO",
    title = "{Final results of Borexino on CNO solar neutrinos}",
    eprint = "2307.14636",
    archivePrefix = "arXiv",
    primaryClass = "hep-ex",
    doi = "10.1103/PhysRevD.108.102005",
    journal = "Phys. Rev. D",
    volume = "108",
    number = "10",
    pages = "102005",
    year = "2023"
}

@article{PhysRevLett.90.021802,
  title = {First Results from KamLAND: Evidence for Reactor Antineutrino Disappearance},
  author = {Eguchi, K. and Enomoto, S. and Furuno, K. and Goldman, J. and Hanada, H. and Ikeda, H. and Ikeda, K. and Inoue, K. and Ishihara, K. and Itoh, W. and Iwamoto, T. and Kawaguchi, T. and Kawashima, T. and Kinoshita, H. and Kishimoto, Y. and Koga, M. and Koseki, Y. and Maeda, T. and Mitsui, T. and Motoki, M. and Nakajima, K. and Nakajima, M. and Nakajima, T. and Ogawa, H. and Owada, K. and Sakabe, T. and Shimizu, I. and Shirai, J. and Suekane, F. and Suzuki, A. and Tada, K. and Tajima, O. and Takayama, T. and Tamae, K. and Watanabe, H. and Busenitz, J. and Djurcic, Z. and McKinny, K. and Mei, D.-M. and Piepke, A. and Yakushev, E. and Berger, B. E. and Chan, Y. D. and Decowski, M. P. and Dwyer, D. A. and Freedman, S. J. and Fu, Y. and Fujikawa, B. K. and Heeger, K. M. and Lesko, K. T. and Luk, K.-B. and Murayama, H. and Nygren, D. R. and Okada, C. E. and Poon, A. W. P. and Steiner, H. M. and Winslow, L. A. and Horton-Smith, G. A. and McKeown, R. D. and Ritter, J. and Tipton, B. and Vogel, P. and Lane, C. E. and Miletic, T. and Gorham, P. W. and Guillian, G. and Learned, J. G. and Maricic, J. and Matsuno, S. and Pakvasa, S. and Dazeley, S. and Hatakeyama, S. and Murakami, M. and Svoboda, R. C. and Dieterle, B. D. and DiMauro, M. and Detwiler, J. and Gratta, G. and Ishii, K. and Tolich, N. and Uchida, Y. and Batygov, M. and Bugg, W. and Cohn, H. and Efremenko, Y. and Kamyshkov, Y. and Kozlov, A. and Nakamura, Y. and Braeckeleer, L. De and Gould, C. R. and Karwowski, H. J. and Markoff, D. M. and Messimore, J. A. and Nakamura, K. and Rohm, R. M. and Tornow, W. and Young, A. R. and Wang, Y.-F.},
  collaboration = {KamLAND Collaboration},
  journal = {Phys. Rev. Lett.},
  volume = {90},
  issue = {2},
  pages = {021802},
  numpages = {6},
  year = {2003},
  month = {Jan},
  publisher = {American Physical Society},
  doi = {10.1103/PhysRevLett.90.021802},
  url = {https://link.aps.org/doi/10.1103/PhysRevLett.90.021802}
}

@article{JUNO:2025gmd,
    author = "Abusleme, Angel and others",
    collaboration = "JUNO",
    title = "{First measurement of reactor neutrino oscillations at JUNO}",
    eprint = "2511.14593",
    archivePrefix = "arXiv",
    primaryClass = "hep-ex",
    month = "11",
    year = "2025"
}

@article{FUKUDA2003418,
title = {The Super-Kamiokande detector},
journal = {Nucl. Instrum. Meth. A},
volume = {501},
number = {2},
pages = {418-462},
year = {2003},
issn = {0168-9002},
doi = {https://doi.org/10.1016/S0168-9002(03)00425-X},
url = {https://www.sciencedirect.com/science/article/pii/S016890020300425X},
author = {S. Fukuda and Y. Fukuda and T. Hayakawa and E. Ichihara and M. Ishitsuka and Y. Itow and T. Kajita and J. Kameda and K. Kaneyuki and S. Kasuga and K. Kobayashi and Y. Kobayashi and Y. Koshio and M. Miura and S. Moriyama and M. Nakahata and S. Nakayama and T. Namba and Y. Obayashi and A. Okada and M. Oketa and K. Okumura and T. Oyabu and N. Sakurai and M. Shiozawa and Y. Suzuki and Y. Takeuchi and T. Toshito and Y. Totsuka and S. Yamada and S. Desai and M. Earl and J.T. Hong and E. Kearns and M. Masuzawa and M.D. Messier and J.L. Stone and L.R. Sulak and C.W. Walter and W. Wang and K. Scholberg and T. Barszczak and D. Casper and D.W. Liu and W. Gajewski and P.G. Halverson and J. Hsu and W.R. Kropp and S. Mine and L.R. Price and F. Reines and M. Smy and H.W. Sobel and M.R. Vagins and K.S. Ganezer and W.E. Keig and R.W. Ellsworth and S. Tasaka and J.W. Flanagan and A. Kibayashi and J.G. Learned and S. Matsuno and V.J. Stenger and Y. Hayato and T. Ishii and A. Ichikawa and J. Kanzaki and T. Kobayashi and T. Maruyama and K. Nakamura and Y. Oyama and A. Sakai and M. Sakuda and O. Sasaki and S. Echigo and T. Iwashita and M. Kohama and A.T. Suzuki and M. Hasegawa and T. Inagaki and I. Kato and H. Maesaka and T. Nakaya and K. Nishikawa and S. Yamamoto and T.J. Haines and B.K. Kim and R. Sanford and R. Svoboda and E. Blaufuss and M.L. Chen and Z. Conner and J.A. Goodman and E. Guillian and G.W. Sullivan and D. Turcan and A. Habig and M. Ackerman and F. Goebel and J. Hill and C.K. Jung and T. Kato and D. Kerr and M. Malek and K. Martens and C. Mauger and C. McGrew and E. Sharkey and B. Viren and C. Yanagisawa and W. Doki and S. Inaba and K. Ito and M. Kirisawa and M. Kitaguchi and C. Mitsuda and K. Miyano and C. Saji and M. Takahata and M. Takahashi and K. Higuchi and Y. Kajiyama and A. Kusano and Y. Nagashima and K. Nitta and M. Takita and T. Yamaguchi and M. Yoshida and H.I. Kim and S.B. Kim and J. Yoo and H. Okazawa and M. Etoh and K. Fujita and Y. Gando and A. Hasegawa and T. Hasegawa and S. Hatakeyama and K. Inoue and K. Ishihara and T. Iwamoto and M. Koga and I. Nishiyama and H. Ogawa and J. Shirai and A. Suzuki and T. Takayama and F. Tsushima and M. Koshiba and Y. Ichikawa and T. Hashimoto and Y. Hatakeyama and M. Koike and T. Horiuchi and M. Nemoto and K. Nishijima and H. Takeda and H. Fujiyasu and T. Futagami and H. Ishino and Y. Kanaya and M. Morii and H. Nishihama and H. Nishimura and T. Suzuki and Y. Watanabe and D. Kielczewska and U. Golebiewska and H.G. Berns and S.B. Boyd and R.A. Doyle and J.S. George and A.L. Stachyra and L.L. Wai and R.J. Wilkes and K.K. Young and H. Kobayashi}
}

@article{BELLERIVE201630,
title = {The Sudbury Neutrino Observatory},
journal = {Nuclear Physics B},
volume = {908},
pages = {30-51},
year = {2016},
note = {Neutrino Oscillations: Celebrating the Nobel Prize in Physics 2015},
issn = {0550-3213},
doi = {https://doi.org/10.1016/j.nuclphysb.2016.04.035},
url = {https://www.sciencedirect.com/science/article/pii/S0550321316300736},
author = {A. Bellerive and J.R. Klein and A.B. McDonald and A.J. Noble and A.W.P. Poon}
}

@article{PATTERSON2013151,
title = {The NOvA experiment: status and outlook},
journal = {Nuclear Physics B - Proceedings Supplements},
volume = {235-236},
pages = {151-157},
year = {2013},
note = {The XXV International Conference on Neutrino Physics and Astrophysics},
issn = {0920-5632},
doi = {https://doi.org/10.1016/j.nuclphysbps.2013.04.005},
url = {https://www.sciencedirect.com/science/article/pii/S0920563213001266},
author = {R.B. Patterson}
}

@article{DUNE:2021tad,
    author = "Hewes, V. and others",
    collaboration = "DUNE",
    title = "{Deep Underground Neutrino Experiment (DUNE) Near Detector Conceptual Design Report}",
    eprint = "2103.13910",
    archivePrefix = "arXiv",
    primaryClass = "physics.ins-det",
    reportNumber = "FERMILAB-PUB-21-067-E-LBNF-PPD-SCD-T",
    doi = "10.3390/instruments5040031",
    journal = "Instruments",
    volume = "5",
    number = "4",
    pages = "31",
    year = "2021"
}

@article{DayaBay:2018yms,
    author = "Adey, D. and others",
    collaboration = "Daya Bay",
    title = "{Measurement of the Electron Antineutrino Oscillation with 1958 Days of Operation at Daya Bay}",
    eprint = "1809.02261",
    archivePrefix = "arXiv",
    primaryClass = "hep-ex",
    doi = "10.1103/PhysRevLett.121.241805",
    journal = "Phys. Rev. Lett.",
    volume = "121",
    number = "24",
    pages = "241805",
    year = "2018"
}

@article{Hyper-Kamiokande:2021frf,
    author = "Abe, K. and others",
    collaboration = "Hyper-Kamiokande",
    title = "{Supernova Model Discrimination with Hyper-Kamiokande}",
    eprint = "2101.05269",
    archivePrefix = "arXiv",
    primaryClass = "astro-ph.IM",
    doi = "10.3847/1538-4357/abf7c4",
    journal = "Astrophys. J.",
    volume = "916",
    number = "1",
    pages = "15",
    year = "2021"
}

@article{hino2019aging,
  author  = {Hino, Y. and Furuta, H. and Suekane, F.},
  title   = {Aging study of {Gd} concentration in {LAB}-based {Gd} loaded liquid scintillator exposed to passivated stainless steel},
  journal = {Journal of Instrumentation},
  volume  = {14},
  number  = {09},
  pages   = {P09007},
  year    = {2019},
  doi     = {10.1088/1748-0221/14/09/P09007}
}

@article{marti2020egads,
  author  = {Marti, Ll. and others},
  title   = {Evaluation of gadolinium's action on water Cherenkov detector systems with {EGADS}},
  journal = {Nucl. Instrum. Meth. A},
  volume  = {959},
  pages   = {163549},
  year    = {2020},
  doi     = {10.1016/j.nima.2020.163549}
}

@article{fukuda2003superk,
  author  = {Fukuda, S. and others},
  title   = {The {Super-Kamiokande} detector},
  journal = {Nucl. Instrum. Meth. A},
  volume  = {501},
  pages   = {418--462},
  year    = {2003},
  doi     = {10.1016/S0168-9002(03)00425-X}
}

@article{XIE2021165459,
title = {A liquid scintillator for a neutrino detector working at --50 degree},
journal = {Nucl. Instrum. Meth. A},
volume = {1009},
pages = {165459},
year = {2021},
issn = {0168-9002},
doi = {https://doi.org/10.1016/j.nima.2021.165459},
url = {https://www.sciencedirect.com/science/article/pii/S0168900221004447},
author = {Zhangquan Xie and Jun Cao and Yayun Ding and Mengchao Liu and Xilei Sun and Wei Wang and Yuguang Xie}
}

@article{Caravaca2020,
  author  = {Caravaca, J. and Land, B. J. and Yeh, M. and Orebi Gann, G. D.},
  title   = {{Characterization of water-based liquid scintillator for Cherenkov and scintillation separation}},
  journal = {Eur. Phys. J. C},
  volume  = {80},
  number  = {9},
  pages   = {867},
  year    = {2020},
  doi     = {10.1140/epjc/s10052-020-8418-4},
  eprint  = {2006.00173},
  archivePrefix = {arXiv},
  primaryClass  = {physics.ins-det}
}

@misc{koshio2026upgradesuperkamiokandegadolinium,
      title={Upgrade of Super-Kamiokande with Gadolinium}, 
      author={Yusuke Koshio and Masayuki Nakahata and Hiroyuki Sekiya and Mark R. Vagins},
      year={2026},
      eprint={2511.03921},
      archivePrefix={arXiv},
      primaryClass={physics.ins-det},
      url={https://arxiv.org/abs/2511.03921}, 
}

\end{document}